\documentclass[apjl]{emulateapj}
\usepackage{amsmath}
\usepackage{comment}
\usepackage{ifthen}

\newcommand{\forloop}[5][1]%
{%
\setcounter{#2}{#3}%
\ifthenelse{#4}%
	{%
	#5%
	\addtocounter{#2}{#1}%
	\forloop[#1]{#2}{\value{#2}}{#4}{#5}%
	}%
	{%
	}%
}%

\newcommand{\ctbd}[1]{}

\newcommand{\lc}{light curve}

\newcommand{\Lc}{Light curve}

\newcommand{\band}[1]{\ensuremath{#1}~band}

\newcommand{\kms}{\ensuremath{\rm km\,s^{-1}}}
\newcommand{\ms}{\ensuremath{\rm m\,s^{-1}}}

\newcommand{\ergscmsq}{\ensuremath{\rm erg\,s^{-1}\,cm^{-2}}}

\newcommand{\vsini}{\ensuremath{v \sin{i}}}
\newcommand{\feh}{\ensuremath{\rm [Fe/H]}}

\newcommand{\rsun}{\ensuremath{R_\sun}}
\newcommand{\msun}{\ensuremath{M_\sun}}
\newcommand{\lsun}{\ensuremath{L_\sun}}

\newcommand{\rstar}{\ensuremath{R_\star}}
\newcommand{\mstar}{\ensuremath{M_\star}}
\newcommand{\lstar}{\ensuremath{L_\star}}

\newcommand{\teffstar}{\ensuremath{T_{\rm eff\star}}}
\newcommand{\rhostar}{\ensuremath{\rho_\star}}
\newcommand{\loggstar}{\ensuremath{\log{g_{\star}}}}

\newcommand{\rpl}{\ensuremath{R_{p}}}
\newcommand{\mpl}{\ensuremath{M_{p}}}

\newcommand{\arstar}{\ensuremath{a/\rstar}}
\newcommand{\zrstar}{\ensuremath{\zeta/\rstar}}

\newcommand{\rjup}{\ensuremath{R_{\rm J}}}
\newcommand{\mjup}{\ensuremath{M_{\rm J}}}

\newcommand{\refsecl}[1]{\mbox{Section \ref{sec:#1}}}

\newcommand{\reftabl}[1]{Table~\ref{tab:#1}}

\newcommand{\flwof}{\mbox{FLWO 1.2\,m}}

\newcommand{\hatcurhtr}{HTR385-001}                                    
\newcommand{\hatcurCCra}{\ensuremath{18^{\mathrm h}18^{\mathrm m}58.32{\mathrm s}}}                                  
\newcommand{\hatcurCCdec}{\ensuremath{+10{\arcdeg}35{\arcmin}50.3{\arcsec}}}                                 
\newcommand{\hatcurCCtwomass}{2MASS~18185842+1035502}                  
\newcommand{\hatcurCCgsc}{GSC~1014-00973}                              
\newcommand{\hatcurCCtassmv}{\ensuremath{10.465\pm0.029}}              
\newcommand{\hatcurCCtassmB}{\ensuremath{10.916\pm0.021}}              
\newcommand{\hatcurCCtassmI}{\ensuremath{10.007\pm0.063}}              
\newcommand{\hatcurCCtassmg}{\ensuremath{10.741\pm0.074}}              
\newcommand{\hatcurCCtassmr}{\ensuremath{10.371\pm0.053}}              
\newcommand{\hatcurCCtassmi}{\ensuremath{10.282\pm0.034}}              
\newcommand{\hatcurCCtwomassJmag}{\ensuremath{9.670\pm0.027}}          
\newcommand{\hatcurCCtwomassHmag}{\ensuremath{9.497\pm0.029}}          
\newcommand{\hatcurCCtwomassKmag}{\ensuremath{9.433\pm0.024}}          
\newcommand{\hatcurLCrprstar}{\ensuremath{0.0968\pm0.0015}}            
\newcommand{\hatcurLCbsq}{\ensuremath{0.031_{-0.023}^{+0.040}}}        
\newcommand{\hatcurLCimp}{\ensuremath{0.177_{-0.084}^{+0.090}}}        
\newcommand{\hatcurLCzeta}{\ensuremath{15.099\pm0.071}}                
\newcommand{\hatcurLCdur}{\ensuremath{0.14578\pm0.00080}}              
\newcommand{\hatcurLCingdur}{\ensuremath{0.01325\pm0.00054}}           
\newcommand{\hatcurLCP}{\ensuremath{2.4652950\pm0.0000032}}            
\newcommand{\hatcurLCPshort}{\ensuremath{2.4653}}                      
\newcommand{\hatcurLCT}{\ensuremath{2455113.48127\pm0.00048}}          
\newcommand{\hatcurSMEiteff}{\ensuremath{6622\pm50}}                   
\newcommand{\hatcurSMEizfeh}{\ensuremath{-0.740\pm0.080}}              
\newcommand{\hatcurSMEizfehshort}{\ensuremath{-0.74}}                  
\newcommand{\hatcurSMEilogg}{\ensuremath{2.79\pm0.10}}                 
\newcommand{\hatcurSMEivsin}{\ensuremath{110.20\pm0.50}}               
\newcommand{\hatcurSMEivmac}{\ensuremath{0.0}}                         
\newcommand{\hatcurSMEivmic}{\ensuremath{0.0}}                         
\newcommand{\hatcurSMEiiteff}{\ensuremath{7500\pm250}}                 
\newcommand{\hatcurSMEiizfeh}{\ensuremath{-0.25\pm0.25}}               
\newcommand{\hatcurSMEiizfehshort}{\ensuremath{-0.25}}                 
\newcommand{\hatcurSMEiilogg}{\ensuremath{4.204\pm0.014}}              
\newcommand{\hatcurSMEiivsin}{\ensuremath{102.8\pm1.1}}                
\newcommand{\hatcurLBiz}{\ensuremath{0.0946}}                          
\newcommand{\hatcurLBiiz}{\ensuremath{0.3459}}                         
\newcommand{\hatcurLBii}{\ensuremath{0.1526}}                          
\newcommand{\hatcurLBiii}{\ensuremath{0.3515}}                         
\newcommand{\hatcurLBig}{\ensuremath{0.3401}}                          
\newcommand{\hatcurLBiig}{\ensuremath{0.3758}}                         
\newcommand{\hatcurLBir}{\ensuremath{0.2150}}                          
\newcommand{\hatcurLBiir}{\ensuremath{0.3764}}                         
\newcommand{\hatcurISOm}{\ensuremath{1.47\pm0.12}}                     
\newcommand{\hatcurISOmlong}{\ensuremath{1.47\pm0.12}}                 
\newcommand{\hatcurISOr}{\ensuremath{1.500\pm0.050}}                   
\newcommand{\hatcurISOrlong}{\ensuremath{1.500\pm0.050}}               
\newcommand{\hatcurISOrho}{\ensuremath{0.615_{-0.036}^{+0.022}}}       
\newcommand{\hatcurISOlogg}{\ensuremath{4.251\pm0.018}}                
\newcommand{\hatcurISOlum}{\ensuremath{6.4\pm1.1}}                     
\newcommand{\hatcurISOmv}{\ensuremath{2.70\pm0.19}}                    
\newcommand{\hatcurISOage}{\ensuremath{1.00_{-0.51}^{+0.67}}}          
\newcommand{\hatcurISOMK}{\ensuremath{2.13\pm0.10}}                    
\newcommand{\hatcurRVKtwosiglim}{\ensuremath{<215.2}}                  
\newcommand{\hatcurRVjitter}{\ensuremath{312\pm70}}                    
\newcommand{\hatcurPPi}{\ensuremath{88.26\pm0.85}}                     
\newcommand{\hatcurPPar}{\ensuremath{5.825_{-0.116}^{+0.069}}}         
\newcommand{\hatcurPParel}{\ensuremath{0.0406\pm0.0011}}               
\newcommand{\hatcurPPmtwosiglim}{\ensuremath{<1.85}}                   
\newcommand{\hatcurPPr}{\ensuremath{1.413\pm0.054}}                    
\newcommand{\hatcurPPrlong}{\ensuremath{1.413\pm0.054}}                
\newcommand{\hatcurPPteff}{\ensuremath{2200\pm76}}                     
\newcommand{\hatcurPPfluxavg}{\ensuremath{5.29\pm0.74}}                
\newcommand{\hatcurPPfluxavgdim}{\ensuremath{9}}                       
\newcommand{\hatcurXAvblendcor}{\ensuremath{0.38\pm0.12}}              
\newcommand{\hatcurXdistredblendcor}{\ensuremath{303\pm13}}            

\newcommand{\hatcurLineProflambdaone}{\ensuremath{37.6^{\circ}}}
\newcommand{\hatcurLineProflambdatwo}{\ensuremath{51.5^{\circ}}}
\newcommand{\hatcurLineProflambdathree}{\ensuremath{-7.7^{\circ}}}
\newcommand{\hatcurLineProflambdalimonesigA}{\ensuremath{-10.9}}
\newcommand{\hatcurLineProflambdalimonesigB}{\ensuremath{-4.1}}
\newcommand{\hatcurLineProflambdalimonesigC}{\ensuremath{34.2}}
\newcommand{\hatcurLineProflambdalimonesigD}{\ensuremath{44.8}}
\newcommand{\hatcurLineProflambdalimonesigE}{\ensuremath{47.3}}
\newcommand{\hatcurLineProflambdalimonesigF}{\ensuremath{55.9}}
\newcommand{\hatcurLineProflambdalimtwosigA}{\ensuremath{-16.7}}
\newcommand{\hatcurLineProflambdalimtwosigB}{\ensuremath{3.3}}
\newcommand{\hatcurLineProflambdalimtwosigC}{\ensuremath{27.6}}
\newcommand{\hatcurLineProflambdalimtwosigD}{\ensuremath{57.4}}
\newcommand{\hatcurLineProfvsini}{\ensuremath{102.1 \pm 1.3}}
\newcommand{\hatcurLineProfuA}{\ensuremath{0.27\pm0.20}}
\newcommand{\hatcurLineProfuB}{\ensuremath{0.798 \pm 0.049}}
\newcommand{\hatcurLineProfuAstd}{\ensuremath{0.84\pm0.15}}
\newcommand{\hatcurLineProfuBstd}{\ensuremath{0.21 \pm 0.17}}
\newcommand{\hatcurLineProfrho}{\ensuremath{26\pm17}}
\newcommand{\hatcurLineProflambdanoGPlimonesigA}{\ensuremath{-12.41}}
\newcommand{\hatcurLineProflambdanoGPlimonesigB}{\ensuremath{-5.78}}
\newcommand{\hatcurLineProflambdanoGPlimonesigC}{\ensuremath{40.89}}
\newcommand{\hatcurLineProflambdanoGPlimonesigD}{\ensuremath{57.32}}
\newcommand{\hatcurLineProflambdanoGPlimtwosigA}{\ensuremath{-13.41}}
\newcommand{\hatcurLineProflambdanoGPlimtwosigB}{\ensuremath{-4.17}}
\newcommand{\hatcurLineProflambdanoGPlimtwosigC}{\ensuremath{39.80}}
\newcommand{\hatcurLineProflambdanoGPlimtwosigD}{\ensuremath{58.23}}
\newcommand{\hatcurLineProfvsininoGP}{\ensuremath{101.46 \pm 0.19}}

\newcommand{\hatcurXCompanionStarMassPARSECA}{\ensuremath{0.61\pm0.10}}
\newcommand{\hatcurXCompanionStarMassPARSECB}{\ensuremath{0.53\pm0.08}}
\newcommand{\hatcurXCompanionOuterSepshort}{\ensuremath{2\farcs7}}
\newcommand{\hatcurXCompanionInnerSepshort}{\ensuremath{0\farcs22}}
\newcommand{\hatcurXCompanionOuterSeplong}{\ensuremath{2\farcs667 \pm 0.001}}
\newcommand{\hatcurXCompanionInnerSeplong}{\ensuremath{0\farcs225 \pm 0.002}}
\newcommand{\hatcurXCompanionOuterPhysSeplong}{\ensuremath{800\pm30}}
\newcommand{\hatcurXCompanionInnerPhysSeplong}{\ensuremath{68\pm3}}
\newcommand{\hatcurXCompanionOuterPeriodApprox}{\ensuremath{14000}}
\newcommand{\hatcurXCompanionInnerPeriodApprox}{\ensuremath{500}}

\newcommand{\hatcur}{HAT-P-57}
\newcommand{\hatcurb}{HAT-P-57b}

\newcommand{\hatcurCCswasp}{1SWASP~J181858.42+103550.1}

\newcommand{\hatcurlumind}{\rhostar}
\newcommand{\hatcurjhkfilset}{ESO}

\newcommand{\hatcurSMEversion}{ii}                                       
\newcommand{\hatcurSMEteff}{\ifthenelse{\equal{\hatcurSMEversion}{i}}{\hatcurSMEiteff}{\hatcurSMEiiteff}}
\newcommand{\hatcurSMEzfeh}{\ifthenelse{\equal{\hatcurSMEversion}{i}}{\hatcurSMEizfeh}{\hatcurSMEiizfeh}}
\newcommand{\hatcurSMEzfehshort}{\ifthenelse{\equal{\hatcurSMEversion}{i}}{\hatcurSMEizfehshort}{\hatcurSMEiizfehshort}}
\newcommand{\hatcurSMElogg}{\ifthenelse{\equal{\hatcurSMEversion}{i}}{\hatcurSMEilogg}{\hatcurSMEiilogg}}
\newcommand{\hatcurSMEvsin}{\ifthenelse{\equal{\hatcurSMEversion}{i}}{\hatcurSMEivsin}{\hatcurSMEiivsin}}
\newcommand{\hatcurSMEvmac}{\ifthenelse{\equal{\hatcurSMEversion}{i}}{\hatcurSMEivmac}{\hatcurSMEiivmac}}
\newcommand{\hatcurSMEvmic}{\ifthenelse{\equal{\hatcurSMEversion}{i}}{\hatcurSMEivmic}{\hatcurSMEiivmic}}

\newcommand{\hatcurSMEteffcirc}{\ifthenelse{\equal{\hatcurSMEversion}{i}}{\hatcurSMEiteffcirc}{\hatcurSMEiiteffcirc}}
\newcommand{\hatcurSMEzfehcirc}{\ifthenelse{\equal{\hatcurSMEversion}{i}}{\hatcurSMEizfehcirc}{\hatcurSMEiizfehcirc}}
\newcommand{\hatcurSMEzfehshortcirc}{\ifthenelse{\equal{\hatcurSMEversion}{i}}{\hatcurSMEizfehshortcirc}{\hatcurSMEiizfehshortcirc}}
\newcommand{\hatcurSMEloggcirc}{\ifthenelse{\equal{\hatcurSMEversion}{i}}{\hatcurSMEiloggcirc}{\hatcurSMEiiloggcirc}}
\newcommand{\hatcurSMEvsincirc}{\ifthenelse{\equal{\hatcurSMEversion}{i}}{\hatcurSMEivsincirc}{\hatcurSMEiivsincirc}}
\newcommand{\hatcurSMEvmaccirc}{\ifthenelse{\equal{\hatcurSMEversion}{i}}{\hatcurSMEivmaccirc}{\hatcurSMEiivmaccirc}}
\newcommand{\hatcurSMEvmiccirc}{\ifthenelse{\equal{\hatcurSMEversion}{i}}{\hatcurSMEivmiccirc}{\hatcurSMEiivmiccirc}}

\newcommand{\hatcurSMEteffeccen}{\ifthenelse{\equal{\hatcurSMEversion}{i}}{\hatcurSMEiteffeccen}{\hatcurSMEiiteffeccen}}
\newcommand{\hatcurSMEzfeheccen}{\ifthenelse{\equal{\hatcurSMEversion}{i}}{\hatcurSMEizfeheccen}{\hatcurSMEiizfeheccen}}
\newcommand{\hatcurSMEzfehshorteccen}{\ifthenelse{\equal{\hatcurSMEversion}{i}}{\hatcurSMEizfehshorteccen}{\hatcurSMEiizfehshorteccen}}
\newcommand{\hatcurSMEloggeccen}{\ifthenelse{\equal{\hatcurSMEversion}{i}}{\hatcurSMEiloggeccen}{\hatcurSMEiiloggeccen}}
\newcommand{\hatcurSMEvsineccen}{\ifthenelse{\equal{\hatcurSMEversion}{i}}{\hatcurSMEivsineccen}{\hatcurSMEiivsineccen}}
\newcommand{\hatcurSMEvmaceccen}{\ifthenelse{\equal{\hatcurSMEversion}{i}}{\hatcurSMEivmaceccen}{\hatcurSMEiivmaceccen}}
\newcommand{\hatcurSMEvmiceccen}{\ifthenelse{\equal{\hatcurSMEversion}{i}}{\hatcurSMEivmiceccen}{\hatcurSMEiivmiceccen}}

\newcounter{planetcounter}

\newboolean{emulateapj}
\setboolean{emulateapj}{true}

\newboolean{rvtablelong}
\setboolean{rvtablelong}{true}

\newboolean{astroph}
\setboolean{astroph}{true}

\shortauthors{Hartman et al.}
\shorttitle{
\hatcur\lowercase{b}
}
\ifthenelse{\boolean{emulateapj}}{
    \newcommand{\titledag}{$\dagger$}
}{
    \newcommand{\titledag}{\dagger}
}

\begin{document}
\title{
\hatcur\lowercase{b}: A Short-Period Giant Planet Transiting A Bright Rapidly Rotating A8V Star Confirmed Via Doppler Tomography
\altaffilmark{\titledag}
}

\author{
    J.~D.~Hartman\altaffilmark{1}, 
    G.~\'A.~Bakos\altaffilmark{1,*,**},
    L.~A.~Buchhave\altaffilmark{2,3},
    G.~Torres\altaffilmark{2},
    D.~W.~Latham\altaffilmark{2},
    G.~Kov\'acs\altaffilmark{4},
    W.~Bhatti\altaffilmark{1},
    Z.~Csubry\altaffilmark{1},
    M.~de~Val-Borro\altaffilmark{1},
    K.~Penev\altaffilmark{1},
    C.~X.~Huang\altaffilmark{1},
    B.~B\'eky\altaffilmark{5},
    A.~Bieryla\altaffilmark{2},
    S.~N.~Quinn\altaffilmark{6},
    A.~W.~Howard\altaffilmark{7},
    G.~W.~Marcy\altaffilmark{8},
    J.~A.~Johnson\altaffilmark{2},
    H.~Isaacson\altaffilmark{8},
    D.~A.~Fischer\altaffilmark{9},    
    R.~W.~Noyes\altaffilmark{2},
    E.~Falco\altaffilmark{2},
    G.~A.~Esquerdo\altaffilmark{2},
    R.~P.~Knox\altaffilmark{10},
    P.~Hinz\altaffilmark{10},
    J.~L\'az\'ar\altaffilmark{11},
    I.~Papp\altaffilmark{11},
    P.~S\'ari\altaffilmark{11}
}

\altaffiltext{1}{Department of Astrophysical Sciences, Princeton
  University, Princeton, NJ 08544, USA; email: jhartman@astro.princeton.edu}
\altaffiltext{$*$}{
Alfred P. Sloan Research Fellow
}
\altaffiltext{$**$}{
Packard Fellow
}
\altaffiltext{2}{Harvard-Smithsonian Center for Astrophysics, Cambridge, MA 02138, USA}
\altaffiltext{3}{Centre for Star and Planet Formation, Natural History Museum of Denmark, University of Copenhagen, DK-1350 Copenhagen, Denmark}
\altaffiltext{4}{Konkoly Observatory, Budapest, Hungary}
\altaffiltext{5}{Google}
\altaffiltext{6}{Department of Physics and Astronomy, Georgia State University, Atlanta, GA 30303, USA}
\altaffiltext{7}{Institute for Astronomy, University of Hawaii, Honolulu, HI 96822, USA}
\altaffiltext{8}{Department of Astronomy, University of California, Berkeley, CA, USA}
\altaffiltext{9}{Department of Astronomy, Yale University, New Haven, CT, USA}
\altaffiltext{10}{Steward Observatory, University of Arizona, 933 N. Cherry Ave., Tucson, AZ 85721, USA}
\altaffiltext{11}{Hungarian Astronomical Association, Budapest, Hungary}
\altaffiltext{$\dagger$}{
Based on observations obtained with the Hungarian-made Automated
Telescope Network. Based in part on observations made with the Keck-I
telescope at Mauna Kea Observatory, HI (Keck time awarded through NASA
programs N029Hr, N108Hr, N154Hr and N130Hr and NOAO programs A289Hr,
and A284Hr). Based in part on observations
made with the Nordic Optical Telescope, operated on the island of La
Palma jointly by Denmark, Finland, Iceland, Norway, and Sweden, in the
Spanish Observatorio del Roque de los Muchachos of the Instituto de
Astrof\'isica de Canarias. Based in part on observations obtained with
the Tillinghast Reflector 1.5\,m telescope and the 1.2\,m telescope,
both operated by the Smithsonian Astrophysical Observatory at the Fred
Lawrence Whipple Observatory in Arizona.
}


\begin{abstract}

\setcounter{footnote}{10}
We present the discovery of \hatcurb{}, a $P=\hatcurLCPshort{}$\,day
transiting planet around a $V=\hatcurCCtassmv$\,mag, $T_{\rm eff} =
\hatcurSMEteff$\,K main sequence A8V star with a projected rotation
velocity of $\vsini = \hatcurLineProfvsini$\,\kms. We measure the
radius of the planet to be $R = \hatcurPPr$\,\rjup\ and, based on RV
observations, place a 95\% confidence upper limit on its mass of $M
\hatcurPPmtwosiglim$\,\mjup. Based on theoretical stellar evolution
models, the host star has a mass and radius of \hatcurISOm\,\msun, and
\hatcurISOr\,\rsun, respectively. Spectroscopic observations made with
Keck-I/HIRES during a partial transit event show the Doppler shadow of
\hatcurb{} moving across the average spectral line profile of
\hatcur{}, confirming the object as a planetary system. We use these
observations, together with analytic formulae that we derive for the
line profile distortions, to determine the projected angle between the
spin axis of \hatcur{} and the orbital axis of \hatcurb{}. The data
permit two possible solutions, with $\hatcurLineProflambdalimtwosigA^{\circ} < \lambda <
\hatcurLineProflambdalimtwosigB^{\circ}$ or
$\hatcurLineProflambdalimtwosigC^{\circ} < \lambda <
\hatcurLineProflambdalimtwosigD^{\circ}$ at 95\% confidence, and with
relative probabilities for the two modes of 26\% and 74\%,
respectively. Adaptive optics imaging with MMT/Clio2 reveals an object
located $\hatcurXCompanionOuterSepshort$ from \hatcur{} consisting of
two point sources separated in turn from each other by
$\hatcurXCompanionInnerSepshort$. The $H$ and $L^{\prime}$-band
magnitudes of the companion stars are consistent with their being
physically associated with \hatcur{}, in which case they are stars of
mass \hatcurXCompanionStarMassPARSECA\,\msun\ and
\hatcurXCompanionStarMassPARSECB\,\msun. \hatcur{} is the most rapidly
rotating star, and only the fourth main sequence A star, known to host
a transiting planet.
\setcounter{footnote}{0}
\end{abstract}

\keywords{
    planetary systems ---
    stars: individual (\hatcur) ---
    techniques: spectroscopic, photometric
}


\section{Introduction}
\label{sec:introduction}

In order to develop a comprehensive understanding of exoplanetary
systems it is necessary to discover and characterize planets around
stars spanning a wide range of masses. Of the 1887 confirmed
exoplanets listed in the NASA Exoplanet
Archive\footnote{\url{http://exoplanetarchive.ipac.caltech.edu},
  accessed 2015 August 27}, only 104 (5.5\%) are around stars with
masses greater than 1.4\,\msun. The majority of these are evolved
sub-giant and giant stars whose planets were discovered through radial
velocity (RV) surveys \citep[the so-called ``Retired A Star'' surveys;
  e.g.,][]{johnson:2011, wittenmyer:2011, sato:2012}, or moderately
evolved stars with planets discovered by photometric transit surveys
(e.g., HAT-P-49, \citealp{bieryla:2014:hat49}, and KELT-7, \citealp{bieryla:2015}, among others).

Finding planets around massive stars via RVs is challenging. The high
surface temperatures of main sequence stars with $M \ga
1.4$\,\msun\ leads to a high ionization fraction for most elements in
their atmospheres and, as a result, their optical spectra have
relatively few deep absorption lines that can be used for precise RV
measurements. Moreover, unlike lower mass stars which lose angular
momentum via magnetized stellar winds, and thus have surface rotation
rates that decrease with increasing age, higher mass stars do not lose
a significant amount of angular momentum over their main sequence
lifetimes, and thus generally rotate at rapid velocities, often
exceeding 100\,\kms\ \citep[e.g.,][]{royer:2007}. The rapid rotation
broadens the absorption lines in the stellar spectra, further reducing
the precision with which their RVs may be measured. For typical A
stars, the combined effects limit the per-point RV precision to a few
100\,\ms\ at best. In order to overcome this limitation, several RV
surveys have targeted evolved stars, thought to have been A stars when
on the main sequence, which have lower surface temperatures and slower
rotation rates compared to their main sequence counterparts.

Results from the Retired A Star RV surveys indicate that these stars
have a population of planets that is significantly different from the
planets around main sequence F, G and K stars
\citep[e.g.,][]{bowler:2010}. In particular, the stars targeted by
these surveys appear to host giant planets significantly more often than F,
G, and K stars, with a markedly different period distribution as
well. This conclusion has not been without controversy, however, with
both the masses of the host stars and the planetary nature of the
detected periodic RV signals being called into question
(\citealp{lloyd:2011,lloyd:2013}, and \citealp{schlaufman:2013}; see,
however, \citealp{johnson:2013} and \citealp{johnsonja:2014} who provide
additional evidence for the retired-A-star nature of the targets). In
any event, one might expect both stellar evolution (leading to the
engulfment and/or evaporation of close-in planets), and the dynamical
evolution of planetary systems through gravitational interactions, to
result in systematic differences in the architectures of planetary
systems around main sequence and post-main-sequence stars.

While there has been some success in finding planets around
post-main-sequence stars with $M > 1.4$\,\msun\ (bearing in mind the
aforementioned caveats), finding planets around massive main sequence
stars remains challenging. To date there are only 14 planets and low
mass brown dwarfs reported around A- or B-type main sequence
stars. Ten of these were discovered by direct imaging (including four
around the A5V star HR~8799, \citealp{marois:2008}, one around the A6V
star $\beta$~Pic, \citealp{lagrange:2009}, two around the A9V star
HIP~73990, \citealp{hinkley:2015}, a candidate around the A7V star
HD~169142, \citealp{biller:2014}, one around the B9V star HIP~78530,
\citealp{lafreniere:2011}, and one around the B9IV star $\kappa$~And,
\citealp{carson:2013}). These objects are on wide separations from
their host stars, and are very massive (in most cases with $M >
10$\,\mjup) and thus represent a substantially different population of
planets from the closer-in and generally lower mass planets discovered
through RVs or the transit technique. Moreover, the masses and radii
of the directly imaged planets are not directly measured, but depend
on theoretical planet evolution and atmosphere models which have large
uncertainties due to the unknown initial conditions set by the planet
formation process \citep[e.g.,][]{spiegel:2012}, and due to the
typically poorly constrained ages of the host stars
\citep[e.g.,][]{baines:2012}.

The other four planets known around A stars were discovered through
the transit technique (WASP-33, \citealp{colliercameron:2010b};
Kepler-13A, \citealp{rowe:2011}, \citealp{shporer:2011:kep13},
\citealp{mazeh:2012}; and KOI-89 with two planets,
\citealp{rowe:2014}). Finding planets around massive stars via
transits is also more challenging than around less massive
stars. Planets of a given size produce shallower transits around
larger stars (though the transit durations are longer around more
massive stars for a given orbital period, which compensates somewhat
for the lower transit depths) making them harder to detect. And, once
detected, it may not be possible to confirm the planets through the RV
detection of the orbital wobble of their host stars for the reasons
already discussed.  None of the previously known transiting exoplanets
(TEPs) around A stars were initially confirmed through the RV
detection of the orbital wobbles of their host stars. Instead,
\citet{colliercameron:2010b} confirmed WASP-33b via Doppler
tomography, Kepler-13Ab was confirmed through the photometric
detection of Doppler beaming and the planet-induced tidal distortion
of its host star, while KOI-89b and KOI-89c were statistically
validated by leveraging the low probability of false positives for
objects with multiple periodic transit signals. RV-based measurements
of the mass of WASP-33b have subsequently been reported by
\citet{kovacs:2013} and \citet{lehmann:2015}, the latter finding a
mass of $2.1\pm0.2$\,\mjup\ based on 248 spectroscopic observations.

While the rapid rotation of main sequence A stars hinders our ability
to confirm and measure the masses of their TEPs through
RV observations, it also presents a unique observational opportunity
to characterize certain properties of these planetary systems. For
very rapidly rotating stars the distortions in the spectral line
profiles produced during planetary transits may be fully resolved
without requiring very high resolution spectrographs, or even
especially stable spectrographs (the motion of the planet shadow is
measured in $\kms$ rather than $\ms$). This allows for a direct
measurement of the track of the planet across the surface of the star
with respect to the projected stellar spin axis. While the projected
angle between the orbit of a TEP and the spin axis of
its host star $\lambda$ may also be measured for slower rotating stars
by detecting the anomalous Doppler shift that results from the line
profile distortions during transit \citep[the Rossiter-McLaughlin
  effect, e.g.,][]{queloz:2001}, fully resolving the line profile
distortions leads to a measurement of this angle that is both
significantly more precise and more accurate.

Rapidly rotating stars are also more oblate than slower rotators,
leading to rapid nodal precession for close-in planets on misaligned
orbits. \citet{johnson:2015} leveraged the precision afforded by
Doppler tomography to measure the change in $\lambda$ and the impact
parameter $b$ of WASP-33b between 2008 and 2014, providing a direct
measurement of the nodal precession rate of the system, and an
observational constraint on the $J_{2}$ gravitational quadrupole
moment of the star. Nodal precession has also been detected for
Kepler-13Ab by \citet{szabo:2012} and by \citet{masuda:2015} who
measured a change in transit duration for this system via {\em Kepler}
photometry. 

The rotation-induced oblateness also leads to a non-uniform surface
brightness profile of the star via the gravity-darkening effect. For
planets on misaligned orbits this produces an asymmetric transit shape
which may be used to measure the true (not projected) angle $\psi$
between the spin axis of the star and the orbital axis of the
planet. This has been done using {\em Kepler} photometry by
\citet{masuda:2015} for Kepler-13Ab and HAT-P-7b, and by
\citet{ahlers:2015} for KOI-89b and KOI-89c.

In this paper we report the discovery of \hatcurb{}, a transiting
giant planet around a rapidly rotating A8V star. With a rotation rate
of $\vsini = \hatcurLineProfvsini$\,\kms, \hatcur{} is the most
rapidly rotating star with a confirmed TEP, and it is also
only the fourth A star with a confirmed TEP. In
Section~\ref{sec:obs} we describe the observations leading to the
discovery and characterization of this planetary system, the data are
analyzed in Section~\ref{sec:analysis}, and we discuss the results in
Section~\ref{sec:discussion}.

\section{Observations}
\label{sec:obs}

\subsection{Photometry}
\label{sec:photometry}

All time-series photometric data that we collected for \hatcur{} are
provided in Table~\ref{tab:phfu}. We discuss these observations below.

\subsubsection{Photometric detection}
\label{sec:detection}

The star \hatcur{} was observed by the HATNet wide-field photometric
instruments \citep{bakos:2004:hatnet} between the nights of UT 2009
May 12 and UT 2009 Sep 14. A total of 622 observations of a $10\fdg 6
\times 10\fdg 6$ field centered at ${\rm R.A.} = 06^{\rm hr}24^{\rm
  min}$, ${\rm Dec.} = +30^{\circ}$ were made with the HAT-5 telescope
in Arizona, and 3202 observations of this same field were made with
the HAT-9 telescope in Hawaii (the count is after filtering
12 outlier measurements). We used a Sloan~$r$ filter and an
exposure time of $300$\,s. Following \citet{bakos:2010:hat11} and
\citet{kovacs:2005:TFA}, we reduced the images to trend-filtered light
curves and searched these for periodic transit signals using the Box-fitting
Least Squares algorithm \citep[BLS;][]{kovacs:2002:BLS}.

\begin{figure}[]
\plotone{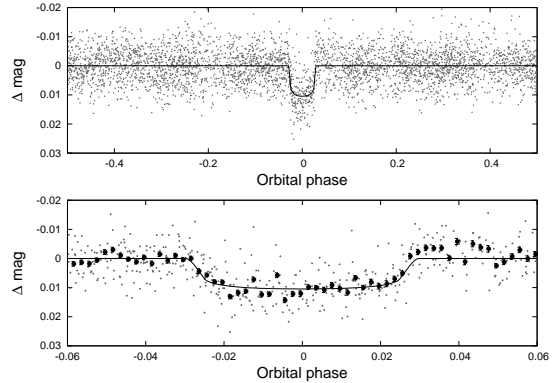}
\caption[]{
    HATNet \lc{} of \hatcur\ phase folded with the transit period.
    The top panel shows the unbinned light curve,
    while the bottom shows the region zoomed-in on the transit, with
    dark filled circles for the light curve binned in phase with a
    bin size of 0.002. The solid line shows the model fit to the light
    curve.
\label{fig:hatnet}}
\end{figure}

Transits were detected in the light curve of \hatcur{} with a period
of $P = \hatcurLCP$\,d. Figure~\ref{fig:hatnet} shows the phase-folded
light curve together with our best-fit model. This same target has
also been included in a list of TEP candidates published
by the Super-WASP survey \citep{lister:2007}, and with a similar
ephemeris, but was shortly thereafter set aside as a probable binary
system based on follow-up spectroscopic observations showing that the
star has a very rapid rotation \citep{colliercameron:2007}. The target
has also been independently identified as a TEP
candidate by the KELT survey (J.~Pepper private communication, July
17, 2015; the KELT project is discussed in \citealp{siverd:2012}).

\begin{figure}[]
\plotone{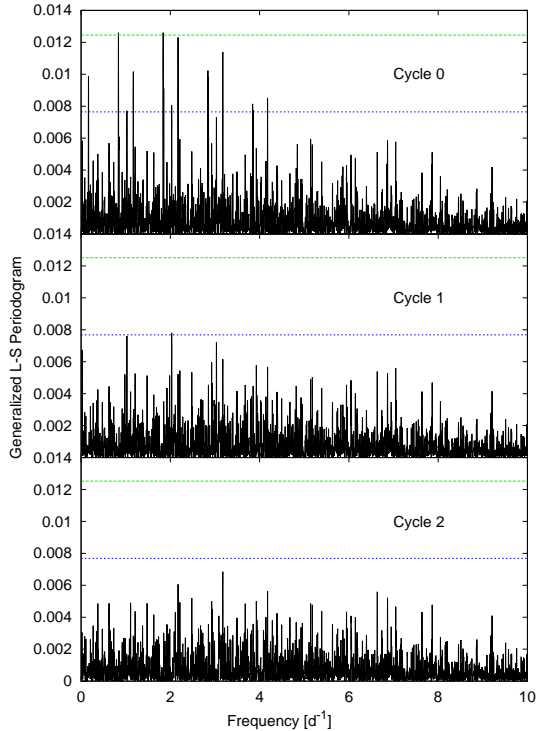}
\caption[]{
    Generalized Lomb-Scargle periodogram of the HATNet \lc{} of
    \hatcur{}. The top panel is the periodogram of the original light
    curve with the transit signal removed, the middle panel is the
    periodogram after whitening the light curve at the highest power
    frequency identified the top periodogram (1.839\,d$^{-1}$ or its
    alias at 0.837\,d$^{-1}$), and the bottom panel is the periodogram
    after whitening the light curve at the highest power frequencies
    identified in the top and middle periodograms (2.030\,d$^{-1}$ or
    its alias at 1.026\,d$^{-1}$). In each case the vertical axis is
    the unnormalized Generalized Lomb-Scargle periodogram given by
    eq.~5 of \citet{zechmeister:2009}. The dotted lines in each panel
    show the periodogram values corresponding to formal false alarm
    probabilities of $10^{-6}$ (upper lined) and $10^{-2}$ (lower
    lined). The periodograms are calculated up to a maximum frequency
    of 100\,d$^{-1}$, but we only display them to a maximum frequency
    of 10\,d$^{-1}$ because all significant power in the light curve
    is found below this frequency.
\label{fig:lsperiodogram}}
\end{figure}

We searched the light curve for additional transit signals or other
periodic variations by running BLS and the Generalized Lomb-Scargle
\citep{zechmeister:2009} algorithms on the residuals from the best-fit
transit model. No additional transit signals were detected, however we
do find two periodic signals at frequencies of 1.839\,d$^{-1}$ (or its
alias at 0.837\,d$^{-1}$) and 2.030\,d$^{-1}$ (or its alias at
1.026\,d$^{-1}$) with false alarm probabilities of $10^{-6.1}$ and
$10^{-2.1}$, respectively, and with peak-to-peak amplitudes of
1.7\,mmag and 1.3\,mmag, respectively. The second frequency is
identified after fitting and subtracting a sinusoid with a frequency
of $1.839$\,d$^{-1}$. A further whitening cycle, with the
2.030\,d$^{-1}$ signal also removed, reveals no other significant
periods in the data (the highest signal in the final periodogram has a
peak-to-peak amplitude of 1.2\,mmag). In each case the aliases are of
comparable significance to the highest peak in the periodogram, and so
we cannot determine whether the true primary frequency is
1.839\,d$^{-1}$ or 0.837\,d$^{-1}$, or whether the true secondary
frequency is 2.030\,d$^{-1}$ or
1.026\,d$^{-1}$. Figure~\ref{fig:lsperiodogram} shows the
periodogram. One or both of these signals may be associated with the
rotation period of the star, or they could correspond to gravity modes
(all frequencies are too low for p-modes), in which case, given its
surface temperature, \hatcur{} would be a $\gamma$ Dor-type variable
(cf., WASP-33 which shows both $\delta$-Scuti and $\gamma$-Dor
variations). The temperature and surface gravity inferred for \hatcur{} in Section~\ref{sec:stellar} place it within both the classical $\delta$-Scuti and $\gamma$-Dor instability strips (e.g., \citealp{rodriguez:2001}, and \citealp{handler:2002}, respectively), so pulsational variability is expected for \hatcur{}.

\subsubsection{Photometric follow-up}
\label{sec:phfu}

\begin{figure}[]
\plotone{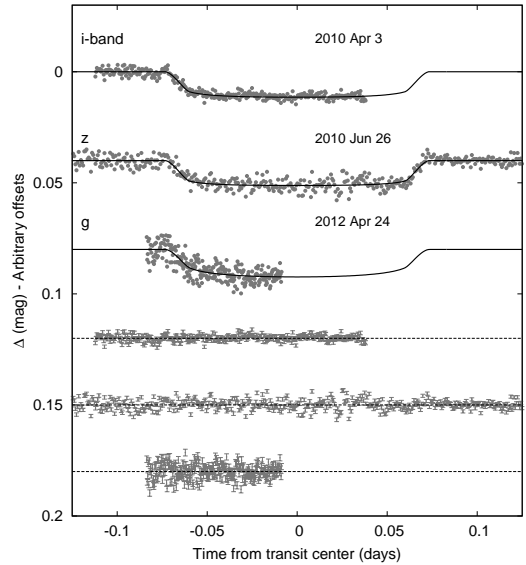}
\caption{
    Unbinned transit light curves for \hatcur, acquired with KeplerCam
    at the \flwof{} telescope. The dates and band-passes are
    indicated. The light curve has been corrected for trends fit
    simultaneously with the transit model.  Our best fit from the
    global modeling described in \refsecl{analysis} is shown by the
    solid line.  Residuals from the fit are displayed below the
    original light curves in the same order.  The error bars represent
    the photon and background shot noise, plus the readout noise.
}
\label{fig:lc}
\end{figure}

Photometric follow-up observations of \hatcur{} were carried out with
KeplerCam on the Fred Lawrence Whipple Observatory (FLWO) 1.2\,m
telescope. We observed ingress events on the nights of 2010 April 3
and 2012 April 24, in $i$ and $g$-bands respectively, and a full
transit on the night of 2010 June 26 in $z$-band. The images were
reduced to light curves following \citet{bakos:2010:hat11}, including
external parameter decorrelation performed simultaneously with the
transit fit to remove systematic trends; the systematics-corrected light
curves are shown in Fig.~\ref{fig:lc}. The r.m.s.~of the residuals from
our best-fit model varies from 1.4\,mmag to 3.4\,mmag for these data.

Additional photometric follow-up observations were carried out with
the FLWO~1.2\,m on the night of 2015 May 12 covering most of a
predicted secondary eclipse event. The observations were performed in
$z$-band and were used to constrain blend scenarios
(Section~\ref{sec:blend}).

\ifthenelse{\boolean{emulateapj}}{
    \begin{deluxetable*}{lrrrrr}
}{
    \begin{deluxetable}{lrrrrr}
}
\tablewidth{0pc}
\tablecaption{
    Differential photometry of
    \hatcur\label{tab:phfu}.
}
\tablehead{
    \colhead{BJD\tablenotemark{a}} & 
    \colhead{Mag\tablenotemark{b}} & 
    \colhead{\ensuremath{\sigma_{\rm Phot}}} &
    \colhead{Mag(orig)\tablenotemark{c}} & 
    \colhead{Filter} &
    \colhead{Instrument} \\
    \colhead{\hbox{~~~~(2,400,000$+$)~~~~}} & 
    \colhead{} & 
    \colhead{} &
    \colhead{} & 
    \colhead{} & 
    \colhead{}
}
\startdata
$ 54998.84656 $ & $   0.00470 $ & $   0.00247 $ & $ \cdots $ & $ r$ &     HATNet\\
$ 55025.96482 $ & $   0.00396 $ & $   0.00220 $ & $ \cdots $ & $ r$ &     HATNet\\
$ 55072.80545 $ & $  -0.00518 $ & $   0.00237 $ & $ \cdots $ & $ r$ &     HATNet\\
$ 55067.87492 $ & $  -0.00243 $ & $   0.00211 $ & $ \cdots $ & $ r$ &     HATNet\\
$ 55021.03497 $ & $  -0.00126 $ & $   0.00251 $ & $ \cdots $ & $ r$ &     HATNet\\
$ 54984.05576 $ & $  -0.00489 $ & $   0.00211 $ & $ \cdots $ & $ r$ &     HATNet\\
$ 55062.94639 $ & $   0.00637 $ & $   0.00208 $ & $ \cdots $ & $ r$ &     HATNet\\
$ 55077.73838 $ & $  -0.00658 $ & $   0.00252 $ & $ \cdots $ & $ r$ &     HATNet\\
$ 55067.87893 $ & $  -0.00097 $ & $   0.00212 $ & $ \cdots $ & $ r$ &     HATNet\\
$ 55025.96892 $ & $   0.00174 $ & $   0.00215 $ & $ \cdots $ & $ r$ &     HATNet\\

\enddata
\tablenotetext{a}{
    Barycentric Julian Date calculated directly from UTC, {\em
      without} correction for leap seconds.
}
\tablenotetext{b}{
    The out-of-transit level has been
    subtracted. These values have been corrected for trends
    simultaneously with the transit fit for the follow-up data. For
    HATNet trends were filtered {\em before} fitting for the transit.
}
\tablenotetext{c}{
    Raw magnitude values after correction using comparison stars, but
    without additional trend-filtering. We only report this value for
    the KeplerCam observations.
}
\tablecomments{
    This table is available in a machine-readable form in the online
    journal.  A portion is shown here for guidance regarding its form
    and content.
}
\ifthenelse{\boolean{emulateapj}}{
    \end{deluxetable*}
}{
    \end{deluxetable}
}

\subsubsection{Adaptive Optics Imaging}
\label{sec:hiresimage}

\begin{figure*}[]
\plottwo{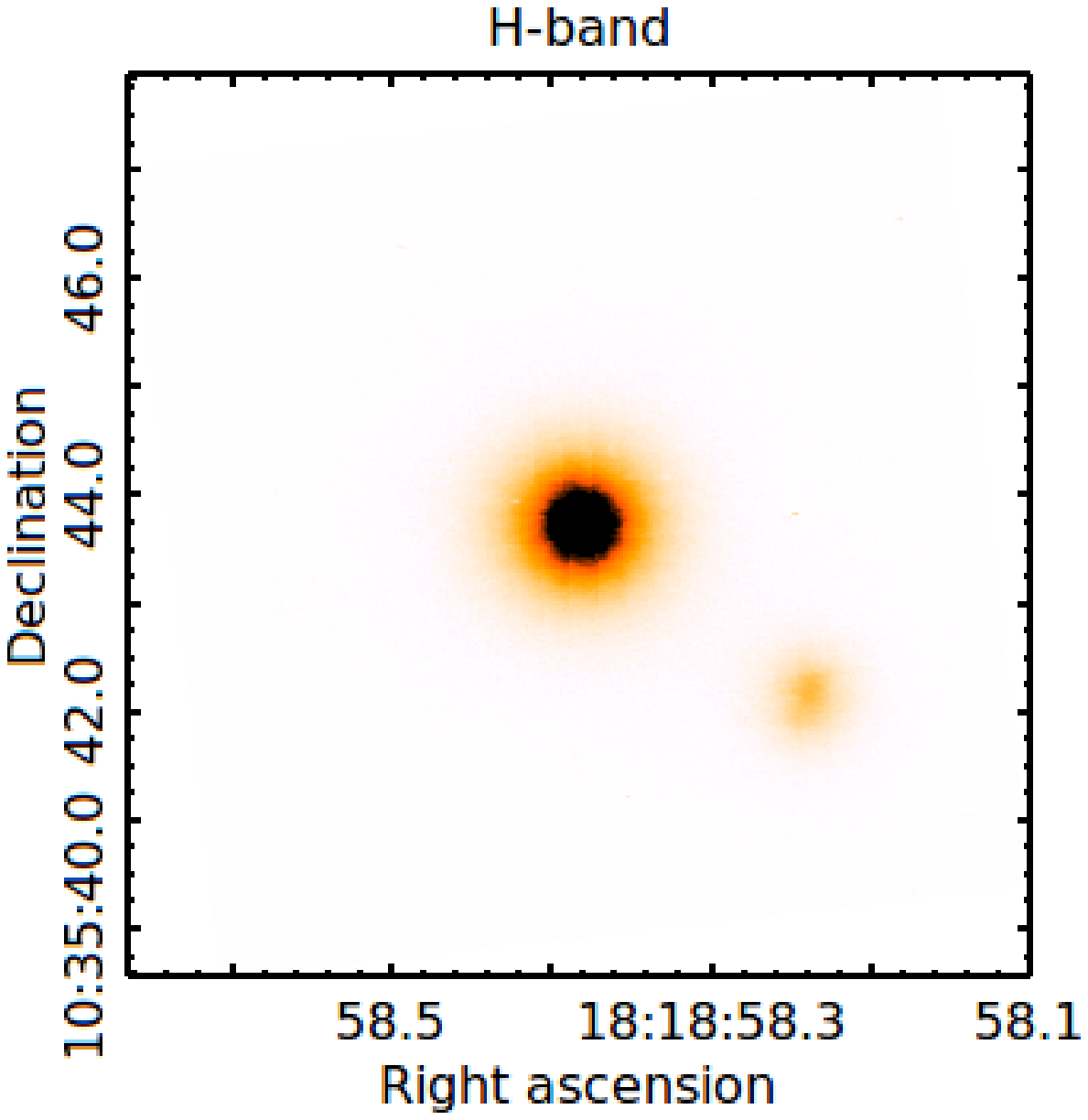}{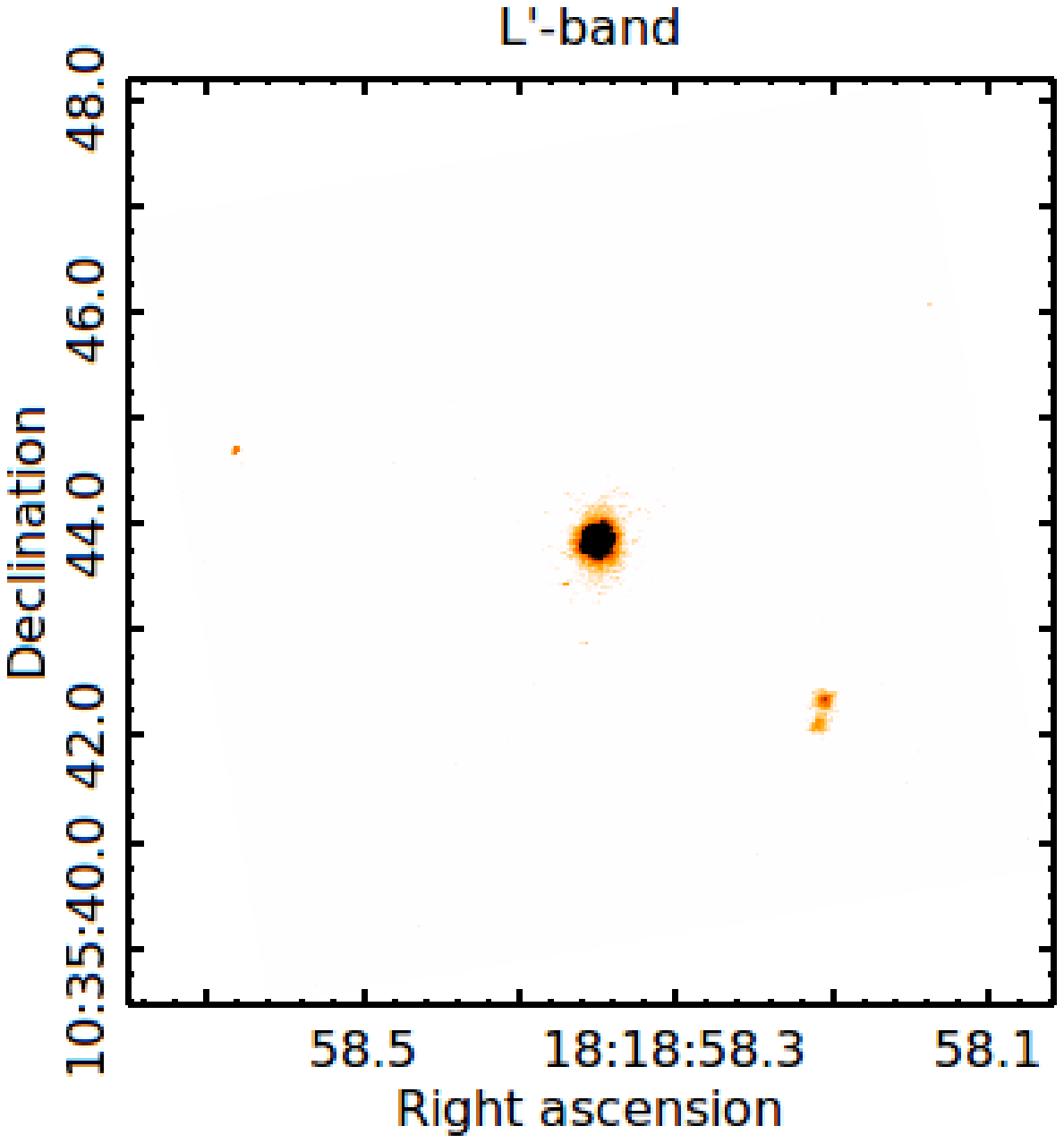}
\caption[]{
    Representative $H$-band (left) and $L^{\prime}$-band (right) MMT/Clio2
    images of \hatcur{}. A faint neighboring source is seen
    $\hatcurXCompanionOuterSepshort$ to the southwest of
    \hatcur{}. The $L^{\prime}$ image shows this source to itself be a
    binary object with two components separated by
    $\hatcurXCompanionInnerSepshort$. Other brightness peaks in the
    $L^{\prime}$ image shown here are either hot pixels or cosmic ray hits.
\label{fig:aoimage}}
\end{figure*}

\begin{figure}[]
\plotone{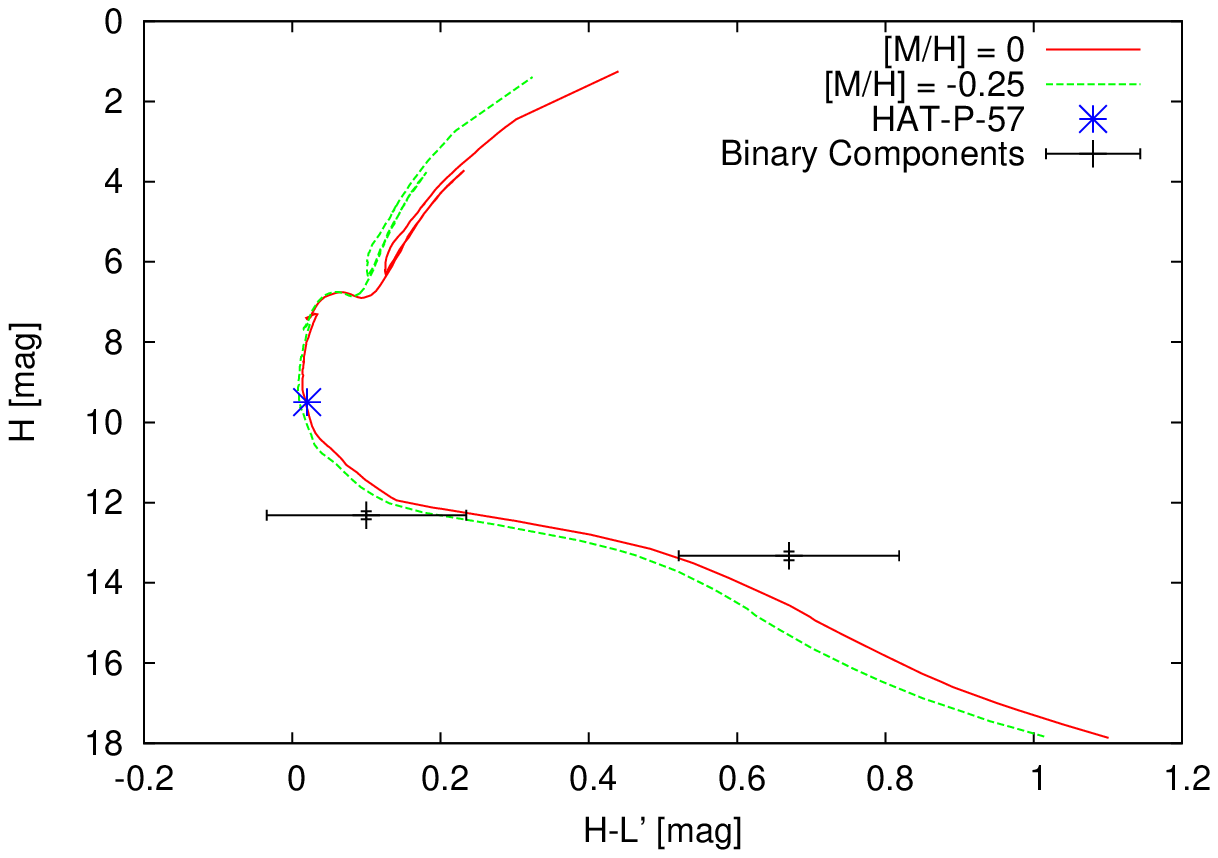}
\caption[]{
  $H-L^{\prime}$ vs.\ $H$ color-magnitude diagram comparing the measured
  apparent magnitudes for \hatcur{} and the components of the nearby
  binary system seen in Figure~\ref{fig:aoimage} to the 1\,Gyr PARSEC
  isochrones \citep{bressan:2012} shifted to the distance of \hatcur{}. The
  $L^{\prime}$ magnitude for \hatcur{} is assumed based on its
  spectral type and measured $H$ magnitude. The components of the
  binary have photometry consistent with the isochrone, placing them
  at the same distance as \hatcur{}, so we conclude that they are
  likely to be physically associated with \hatcur{}.
\label{fig:isoccompbinary}}
\end{figure}

We obtained high-resolution imaging of \hatcur{} on the night of UT
2011 June 22 using the Clio2 near-IR imager \citep{freed:2004} on the MMT
6.5\,m telescope on Mt.~Hopkins, in AZ. Observations in $H$-band and
$L^{\prime}$-band were made using the adaptive optics (AO)
system. Figure~\ref{fig:aoimage} shows the resulting images which
reveal the presence of a binary pair of stars located $\hatcurXCompanionOuterSeplong$ from
\hatcur{}. The pair of stars is resolved into a $\hatcurXCompanionInnerSeplong$ binary in
the $L^{\prime}$ image. In $H$-band the two objects are not cleanly
resolved, but the PSF is clearly elongated. Applying aperture
photometry to the $L^{\prime}$ observations we find that the two
components have $\Delta L^{\prime}$ magnitudes relative to \hatcur{}
of $\Delta L^{\prime}_{B} = 2.72 \pm 0.09$\,mag and $\Delta
L^{\prime}_{C} = 3.16 \pm 0.10$\,mag, respectively. To determine the
relative $H$-band magnitudes we perform PSF fitting fixing the
relative positions of the binary components to those from the
$L^{\prime}$ images and using \hatcur{} to define the PSF. We find
$\Delta H_{B} = 2.82 \pm 0.10$\,mag and $\Delta H_{C} = 3.83 \pm
0.11$\,mag. 

Figure~\ref{fig:isoccompbinary} shows the location of the two binary
components, together with \hatcur{}, on a $H-L^{\prime}$ vs.\ $H$
CMD. We also show 1.0\,Gyr isochrones from the PARSEC model
\citep{bressan:2012} with metallicities of [M/H]$=-0.25$ and [M/H]$=0.0$, and
shifted to the distance of \hatcur{} inferred in
Section~\ref{sec:stellar}. We show the PARSEC model isochrones, rather
than the Y$^{2}$ isochrones which are used in Section~\ref{sec:stellar} to
determine the physical parameters of \hatcur{}, because the PARSEC
models provide a better match to the NIR photometry of M dwarf
stars. The three stars are consistent with being on the same
isochrone, so we conclude that the binary objects are likely to be
physically associated with \hatcur{}. Assuming this is the case, we adopt the names \hatcur{}B and \hatcur{}C for the components of the binary objects, and estimate their masses to be
\hatcurXCompanionStarMassPARSECA\,\msun\ and
\hatcurXCompanionStarMassPARSECB\,\msun, respectively. The $\hatcurXCompanionInnerSeplong$
angular separation between \hatcur{}B and \hatcur{}C corresponds to a
projected physical separation of $\hatcurXCompanionInnerPhysSeplong$\,AU, and approximate orbital period
of $\hatcurXCompanionInnerPeriodApprox$\,yr (assuming the projected separation corresponds to the physical
semimajor axis of the orbit), while the $\hatcurXCompanionOuterSeplong$ separation
between the binary objects and \hatcur{} corresponds to a projected
physical separation of $\hatcurXCompanionOuterPhysSeplong$\,AU and approximate orbital period of $\hatcurXCompanionOuterPeriodApprox$\,yr.

Note that although we do not spatially resolve the binary object from
\hatcur{} in any of our photometric light curves, the Doppler
tomography observations prove that the transiting component is
orbiting the bright A star rather than either of the fainter
components (Section~\ref{sec:lineprof}). We also note that even
without the Doppler tomography observations we would still be able to
draw this conclusion as the binary object is too faint (in the
optical), and its components are too red, to be responsible for the
1\% transits seen consistently with KeplerCam in the $g$, $i$ and
$z$-bands. Given the $H-L^{\prime}$ colors of the binary objects, we
expect \hatcur{}B to be $6.2$\,mag fainter than \hatcur{} in $g$-band,
$4.4$\,mag fainter than \hatcur{} in $i$-band, and $4.0$\,mag fainter
in $z$-band.  Even if \hatcur{}B were totally eclipsed, the blended
eclipse depth of 0.3\% in $g$ would be too shallow to produce the
observed 1\% deep transits. Moreover, transits in $i$ and $z$ would be
significantly deeper than in $g$, which is not what we
observe. 

\begin{figure}[]
\plotone{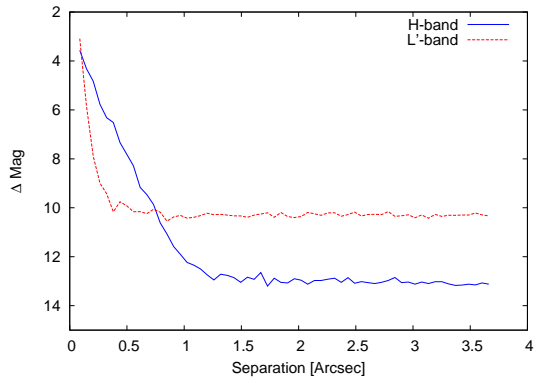}
\caption[]{
  $5\sigma$ detection limit for any additional companions to \hatcur{}
  as a function of angular separation from \hatcur{}. These curves are based on the
  observations described in Section~\ref{sec:hiresimage}.
\label{fig:contrastcurve}}
\end{figure}

Figure~\ref{fig:contrastcurve} shows the approximate 5$\sigma$
detection limits for any additional companions to \hatcur{} in the $H$
and $L^{\prime}$-bands as a function of angular separation. These are
estimated as five times the standard deviation of the pixel values in
circular annuli centered on \hatcur{}, relative to the peak pixel value
of \hatcur{}.

\subsection{Spectroscopy}
\label{sec:hispec}

We carried out spectroscopic observations of \hatcur{} between UT 2010
April 5 and UT 2010 July 1 with the Tillinghast Reflector Echelle
Spectrograph \citep[TRES;][]{furesz:2008} on the 1.5\,m Tillinghast
Reflector at FLWO. We also obtained spectra of \hatcur{} on UT 2010
July 1--3 with the Fibre-fed Echelle Spectrograph
\citep[FIES;][]{telting:2014} on the 2.56\,m Nordic Optical Telescope
(NOT) at the Observatorio del Roque de los Muchachos, on La Palma,
Spain. Additional spectra were obtained using HIRES \citep{vogt:1994}
on the Keck-1 10\,m telescope between UT 2010 June 27 and UT 2012
March 10. A total of 24 HIRES observations were collected during this
time period, including 14 observations made through the I$_{2}$ cell
\citep[e.g.][]{marcy:1992}, and 10 observations without the I$_{2}$
cell. These latter observations were obtained on the night of UT 2010
June 27, primarily during a planetary transit
(Section~\ref{sec:lineprof} discusses the analysis of these
observations in more detail).

The TRES and FIES observations were reduced to initial RVs, bisector
spans (BSs) and stellar atmospheric parameters following
\cite{buchhave:2010:hat16}. Higher precision stellar atmospheric
parameters were also measured from these observations using the
Stellar Parameter Classification program
\citep[SPC;][]{buchhave:2012:spc}. These measurements clearly indicate
that \hatcur{} is a rapidly rotating star with $v\sin i \ga 100$\,\kms,
and a surface temperature hotter than $7000$\,K. Due to the very broad
absorption lines, however, the surface gravity cannot be reliably determined
from the spectra with these techniques. Finally we note that the
spectra do not appear to be composite, and also show no large RV
variations. The five TRES RVs have an r.m.s.\ scatter of 2.2\,\kms,
while the three FIES RVs have an r.m.s.\ scatter of 0.35\,\kms. The difference in precision is largely due to using a single spectral order to measure the TRES RVs, compared to five orders used for FIES (a multi-order analysis of the TRES data would yield more precise measurements, but given the lower S/N of the spectra compared to those from FIES, we expect the scatter would still exceed that of the FIES RVs).

\begin{figure} []
\plotone{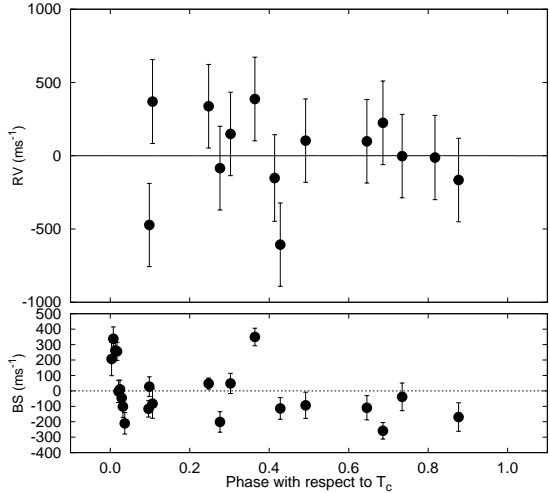}
\caption{
    {\em Top panel:} RV measurements from Keck-I/HIRES computed using the I$_{2}$-Doppler method, and shown as a
    function of orbital phase.  Zero phase corresponds to the time
    of mid-transit.  The center-of-mass velocity has been subtracted. The error bars include a ``jitter'' component
    (\hatcurRVjitter\,\ms) added in quadrature to the formal errors. Due to the large scatter in the velocities resulting from the rapid rotation of the host star we do not detect the orbital variation of the star due to the planet. Based on these observations we place a 95\% confidence upper limit on the orbital semi-amplitude of $K\hatcurRVKtwosiglim$\,\ms.
    {\em Bottom panel:} Bisector spans (BS). These are shown for the I$_{2}$-free observations as well as the observations taken with the I$_{2}$-cell in. A zoom-in on the in-transit measurements is shown in Figure~\ref{fig:rvbsrm}.
    Note the different vertical scales of the panels.
}
\label{fig:rvbis}
\end{figure}

Wavelength calibrated spectra were extracted from the HIRES echelle
images using the reduction pipeline of the California Planet Search
team. The 14 I$_{2}$-in observations were reduced to relative RVs in
the barycentric frame of the Solar System following the method of
\citet{butler:1996}. For this we made use of the highest S/N
out-of-transit I$_{2}$-free observation as a template.  These are
shown phase-folded with the orbital ephemeris in
Figure~\ref{fig:rvbis}. We also measured spectral line bisector spans
(BSs) from the I$_{2}$-free blue orders for 22 of the observations
following \citet{torres:2007:hat3}. Wavelength extracted spectra were
not available for two observations and were excluded from the BS
analysis. The BS values are also shown in Figure~\ref{fig:rvbis}. Our
procedure for measuring the BSs involves cross-correlating the
observed spectra against a synthetic template with atmospheric
parameters similar to those measured for \hatcur{}. We used these same
cross-correlations to measure the barycentric-corrected RVs for the
spectra, including the 10 I$_{2}$-free observations made on UT 2010
June 27. Due to the slit-fed nature of HIRES, and the lack of a
simultaneously obtained wavelength calibration reference, the RV
precision from this CCF method is substantially lower than the
precision obtained from the standard I$_{2}$ Doppler pipeline. More
precise CCF-based RVs were also measured from the 10 I$_{2}$-free
observations using both the blue and green spectral
orders. Table~\ref{tab:rvs} gives the relative RV measurements
obtained with the I$_{2}$ Doppler pipeline, the RV measurements
obtained from the CCFs, and the BSs for the HIRES observations.

\begin{figure} []
\plotone{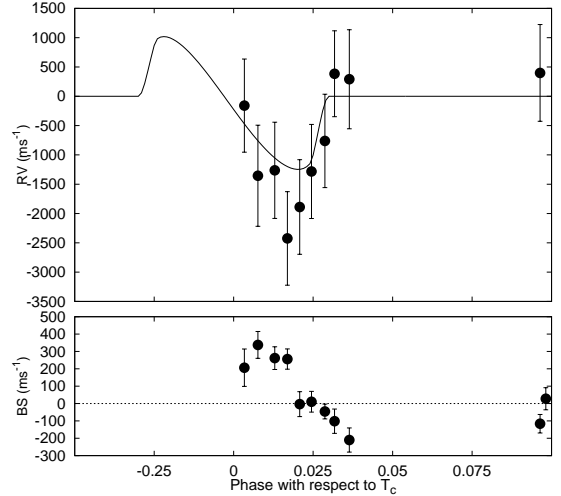}
\caption{
    {\em Top panel:} CCF-based RV measurements from the I$_{2}$-free
    Keck-I/HIRES observations made during (first 7 observations) and
    just after transit (last three observations), shown as a function
    of orbital phase. The Rossiter-McLaughlin effect with an amplitude
    of $\sim 2$\,\kms\ is seen during transit. For reference we also
    show the predicted RV variation from the ARoME model
    \citep{boue:2013} for the maximum posterior probability solution
    to the spectral line profiles
    (Figure~\ref{fig:broadeningprofile}). The model underpredicts the
    measured variation, but since the star is rotating much faster
    than the maximum velocity at which the model is applicable, such a
    discrepancy may be expected.
    {\em Bottom panel:} The BSs (computed for I$_{2}$-in observations
    as well as the I$_{2}$-free observations) also show an anomalous
    variation during transit due to the spectral line profile
    distortion caused by the TEP.
}
\label{fig:rvbsrm}
\end{figure}

The CCF-based RVs and the BSs measured from the in-transit HIRES
observations both show evidence of the Rossiter-McLaughlin effect
(Figure~\ref{fig:rvbsrm}). Note that the sense of the variation seen in both indicators is consistent. For the BS values plotted, we are using the definition
\begin{equation}
BS = RV_{\rm CCF, min} - RV_{\rm CCF, max}
\end{equation}
where $RV_{\rm CCF, min}$ is the bisector velocity at the low CCF
value (i.e., in the wings of the absorption line), while $RV_{\rm CCF,
  max}$ is the bisector velocity at the high CCF value (in the core of
the line). A positive BS indicates that the core of the line is
blue-shifted compared to the wings of the line. Thus with this
definition, we expect the BS and RV variations to be
anti-correlated. Rather than attempting to measure the projected
spin--orbit alignment angle $\lambda$ from these observations we
perform a Doppler tomography analysis of the line profile distortions
in Section~\ref{sec:lineprof}. In Figure~\ref{fig:rvbsrm} we also show
the approximate expected RV variation due to the RM effect calculated using the
ARoME package \citep{boue:2013} for the maximum posterior probability
solution determined from the line profile modeling. This model
underpredicts the anomalous Doppler shift, and if we attempt to fit
the model directly to the observations then the results require a very
high value for $v \sin i$ (175\,\kms), which is completely
inconsistent with the width of the line profiles seen in the HIRES
spectra. The model, however, is not applicable to very high rotation
rates ($v \sin i > 20$\,\kms), and so such a discrepancy is not
unexpected.

\setcounter{planetcounter}{1}
%

\ifthenelse{\boolean{emulateapj}}{
    \begin{deluxetable*}{lrrrrrrrrr}
}{
    \begin{deluxetable}{lrrrrrrrrr}
}
\tablewidth{0pc}
\tablecaption{
    Radial velocities, and bisector span measurements of \hatcur\ from Keck-I/HIRES.
    \label{tab:rvs}
}
\tablehead{
    \colhead{BJD\tablenotemark{a}} &
    \colhead{RV I${_2}$ \tablenotemark{b}} &
    \colhead{\ensuremath{\sigma_{\rm RV I_{2}}}\tablenotemark{c}} &
    \colhead{RV CCF B \tablenotemark{d}} &
    \colhead{\ensuremath{\sigma_{\rm RV CCF B}}} &
    \colhead{RV CCF B+G \tablenotemark{e}} &
    \colhead{\ensuremath{\sigma_{\rm RV CCF B+G}}} &
    \colhead{BS} &
    \colhead{\ensuremath{\sigma_{\rm BS}}} &
    \colhead{Phase \tablenotemark{f}} \\
    \colhead{\hbox{(2,455,000$+$)}} &
    \colhead{(\ms)} &
    \colhead{(\ms)} &
    \colhead{(\kms)} &
    \colhead{(\kms)} &
    \colhead{(\kms)} &
    \colhead{(\kms)} &
    \colhead{(\ms)} &
    \colhead{(\ms)} &
    \colhead{}
}
\startdata
374.81081 & $ \cdots $ & $ \cdots $ & -8.45 & 0.97 & -7.30 & 0.74 & 206 & 108 & 0.003 \\
374.82136 & $ \cdots $ & $ \cdots $ & -9.75 & 1.02 & -8.49 & 0.81 & 338 & 77 & 0.008 \\
374.83442 & $ \cdots $ & $ \cdots $ & -9.49 & 0.99 & -8.40 & 0.77 & 261 & 66 & 0.013 \\
374.84418 & $ \cdots $ & $ \cdots $ & -10.87 & 0.86 & -9.56 & 0.74 & 256 & 58 & 0.017 \\
374.85378 & $ \cdots $ & $ \cdots $ & -10.31 & 0.91 & -9.03 & 0.75 & -3 & 72 & 0.021 \\
374.86293 & $ \cdots $ & $ \cdots $ & -9.41 & 0.97 & -8.42 & 0.75 & 10 & 60 & 0.024 \\
374.87329 & $ \cdots $ & $ \cdots $ & -8.92 & 0.97 & -7.90 & 0.74 & -46 & 43 & 0.029 \\
374.88075 & $ \cdots $ & $ \cdots $ & -7.84 & 0.83 & -6.75 & 0.67 & -102 & 70 & 0.032 \\
374.89223 & $ \cdots $ & $ \cdots $ & -8.13 & 0.95 & -6.85 & 0.79 & -210 & 70 & 0.036 \\
375.04019 & $ \cdots $ & $ \cdots $ & -7.45 & 1.09 & -6.74 & 0.77 & -116 & 53 & 0.096 \\
375.04462 & -492 & 45 & -7.99 & 1.08 & $ \cdots $ & $ \cdots $ & 28 & 64 & 0.098 \\
375.85754 & -626 & 48 & -8.68 & 0.99 & $ \cdots $ & $ \cdots $ & -114 & 70 & 0.428 \\
467.82924 & -22 & 49 & -8.66 & 1.00 & $ \cdots $ & $ \cdots $ & -39 & 90 & 0.734 \\
607.15061 & 319 & 52 & -9.34 & 1.00 & $ \cdots $ & $ \cdots $ & 48 & 36 & 0.247 \\
608.13252 & 80 & 53 & -9.03 & 1.06 & $ \cdots $ & $ \cdots $ & -109 & 79 & 0.646 \\
612.15141 & -104 & 57 & -7.94 & 0.90 & $ \cdots $ & $ \cdots $ & -201 & 66 & 0.276 \\
613.16261 & 206 & 54 & -7.09 & 1.01 & $ \cdots $ & $ \cdots $ & -259 & 53 & 0.686 \\
699.91700 & -185 & 51 & -7.66 & 0.93 & $ \cdots $ & $ \cdots $ & -169 & 93 & 0.876 \\
701.11852 & 368 & 54 & -9.37 & 0.96 & $ \cdots $ & $ \cdots $ & 350 & 57 & 0.364 \\
703.89871 & 84 & 48 & -9.36 & 0.86 & $ \cdots $ & $ \cdots $ & -93 & 85 & 0.491 \\
705.89808 & 130 & 50 & -9.17 & 0.86 & $ \cdots $ & $ \cdots $ & 49 & 65 & 0.302 \\
707.87969 & 351 & 58 & -6.93 & 1.04 & $ \cdots $ & $ \cdots $ & -83 & 95 & 0.106 \\
879.73744 & -31 & 64 & $ \cdots $ & $ \cdots $ & $ \cdots $ & $ \cdots $ & $ \cdots $ & $ \cdots $ & 0.817 \\
997.07656 & -171 & 95 & $ \cdots $ & $ \cdots $ & $ \cdots $ & $ \cdots $ & $ \cdots $ & $ \cdots $ & 0.413 \\

\enddata
\tablenotetext{a}{
    Barycentric Julian Date calculated directly from UTC, {\em
      without} correction for leap seconds.
}
\tablenotetext{b}{
    RVs computed using the I$_{2}$ method. The zero-point of these
    velocities is arbitrary. An overall offset fitted to these
    velocities in \refsecl{analysis} has {\em not} been
    subtracted. Spectra obtained without the I$_{2}$-cell in do not
    have an RV measurement listed in this column.
}
\tablenotetext{c}{
    Internal errors excluding the component of astrophysical jitter
    considered in \refsecl{analysis}.
}
\tablenotetext{d}{
    RVs computed using the CCF method, applied only to the
    I$_{2}$-free blue spectral orders. Note that the units here are
    $\kms$ rather than $\ms$.
}
\tablenotetext{e}{
    RVs computed using the CCF method, applied to the blue and green
    spectral orders. Observations obtained with the I$_{2}$-cell in do not have
    a measurement listed here.
}
\tablenotetext{f}{
    Orbital phase, with phase zero corresponding to mid-transit.
}
\ifthenelse{\boolean{rvtablelong}}{
}{
} 
\ifthenelse{\boolean{emulateapj}}{
    \end{deluxetable*}
}{
    \end{deluxetable}
}

\section{Analysis}
\label{sec:analysis}

\subsection{Stellar Parameters}
\label{sec:stellar}

We measured the stellar atmospheric parameters for \hatcur{} in two
ways. First we analyzed both the HIRES I$_{2}$-free observations and
the FIES observations with SPC. For the second method we performed a
$\chi^2$ comparison of synthetic templates from the Pollux database
\citep{palacios:2010} to the HIRES observations.

The two HIRES orders covered by the SPC library yield substantially
different results for the temperature, metallicity and surface
gravity. For one of the orders we find $T_{\rm eff} = 6620$\,K, $\log
g = 3.39$, [M/H]$= -1.08$ and $v \sin i = 101.7$\,\kms. The normalized
CCF has a peak height of $0.978$ indicating a good match between the
observations and the synthetic template.  For the other order we find
$T_{\rm eff} = 8450$\,K, $\log g = 4.37$, [M/H]$= 0.01$ and $v \sin i
= 102.3$\,\kms. The CCF in this case has a peak height of $0.984$,
again indicating a good match. When the surface gravity is fixed to
$4.20$, based on the Y$^{2}$ stellar evolution models, and using the transit-based stellar density and the effective temperature and metallicities estimated from the spectra (Section~\ref{sec:globmod}; note that we find that the surface gravity from this analysis is
quite well constrained despite the very large initial uncertainty in
the temperature and metallicity), we find temperatures of $T_{\rm eff}
= 7250$\,K and $T_{\rm eff} = 8280$\,K, and metallicities of [M/H]$=
-0.70$ and $=-0.03$ from the respective orders.  Due to the lack of
consistency between the two orders, we conclude that the spectral
overlap between HIRES and the SPC library is too small, in this case,
for a reliable determination of the atmospheric parameters.

When we analyzed the three FIES observations with SPC we find $T_{\rm
  eff} = 6830 \pm 120$\,K, $\log g = 2.95 \pm 0.15$, [M/H]$=-0.82 \pm
0.04$, and $v \sin i = 103.7 \pm 0.9$\,\kms, with a cross-correlation
peak-height of 0.914. The uncertainties are the standard deviation of
the measurements from the three separate observations. When the
surface gravity is fixed to $\log g = 4.20$, we find $T_{\rm eff} =
7440 \pm 80$\,K, [M/H]$ = -0.39 \pm 0.05$, $v \sin i = 102.8 \pm
1.1$\,\kms, and a cross-correlation peak-height of 0.910. The listed
uncertainties reflect the precision, but not the accuracy, of the
measurements. In particular they do not account for the degeneracies
between the parameters, which are more significant at such high
rotation velocities than they are for slower rotating stars where the
true SPC errors have been well calibrated.

For an A star like \hatcur{} the wavelength range of 5050\,\AA\ to
5360\,\AA\ used by SPC does not contain very many good absorption
lines for determining the atmospheric parameters, resulting in
significant degeneracies between the parameters. We therefore carried
out a separate analysis of the Keck/HIRES observations of \hatcur{},
focusing in this case on 18 blue orders covering the wavelength range
3840\,\AA\ to 4793\,\AA. This range of the spectrum contains several
Hydrogen Balmer lines, whose broad profiles constrain the temperature,
as well as many ionized metal lines which are useful for determining
both the metallicity and the temperature.

We first normalized the continua of the observed spectra by fitting
polynomials in wavelength and order to the numerous continuum regions
available for this rapidly rotating star. We then used the Pollux
database \citep{palacios:2010} to obtain a grid of synthetic high
resolution spectra generated using the MARCS atmosphere models
\citep{gustafsson:2008}. The grid spans the temperature range 6500\,K
to 8000\,K in 250\,K steps, metallicities from [M/H]$= -1.0$ to $1.0$
in 0.25\,dex steps, and surface gravities of $\log g = 4.0$ and $\log
g = 4.5$. We applied a rotational broadening kernel with $v \sin i =
102.8$\,\kms, based on the SPC analysis of the FIES data, and assuming
a linear limb darkening law with a coefficient of 0.6, to the
templates and also applied the same continuum normalization procedure
as performed on the observed spectra. Working order by order, we then
cross-correlated each template against the observed spectrum to
determine the red-shift, applied the red-shift to the template,
interpolated the red-shifted template to the wavelength grid of the
observations, and measured the $\chi^2$ difference between the
template and observations ignoring the edges of the order where the
errors are high and the blaze-function and wavelength solution have
systematic errors (the wavelength range to use was determined manually
for each order). For each $\log g$ value we fit a polynomial to the
$\chi^2$-[M/H]-$T_{\rm eff}$ surface to determine the $T_{\rm eff}$
and [M/H] values which minimize the total $\chi^2$ for a spectrum. For
$\log g = 4.0$ we find $T_{\rm eff} = 7476 \pm 11$\,K and [M/H]$ =
-0.2589 \pm 0.0037$, while for $\log g = 4.5$ we find $T_{\rm eff} =
7541 \pm 10$\,K and [M/H]$ = -0.2337 \pm 0.0039$. The errors here are
the standard deviation of the results, demonstrating that this
procedure yields very consistent parameters from observation to
observation. The real errors, however, are dominated by systematic
errors in the models and the relatively low temperature and
metallicity resolution of the grid. For simplicity we adopt the grid
resolution for our estimated errors, with $T_{\rm eff} = 7500 \pm
250$\,K, and [M/H]$ = -0.25 \pm 0.25$. These results are consistent
with the parameters estimated from the FIES spectra with SPC, and are
the parameters that we adopt for the remainder of this paper.

We note that in appearance the spectra are consistent with a late A or
early F classification. Based on the temperature--spectral type scale
from \citet{pecaut:2013}, a temperature of $T_{\rm eff} = 7500$\,K
corresponds to a spectral type of A8.

\begin{figure}[]
\plotone{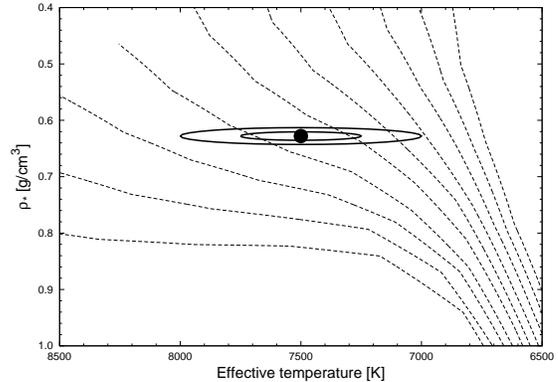}
\caption[]{
    Comparison between the measured values of \teffstar\ and
    \rhostar\ (filled circle), and the Y$^{2}$ model isochrones from
    \citet{yi:2001}. The best-fit values, and approximate 1$\sigma$
    and 2$\sigma$ confidence ellipsoids are shown. The Y$^{2}$
    isochrones are shown for ages of 0.2 to 2.0\,Gyr, in
    0.2\,Gyr increments.
\label{fig:iso}}
\end{figure}

The adopted values for $T_{\rm eff}$ and [M/H], together with the
transit-based mean stellar density (Section~\ref{sec:globmod}), were
then combined with the Yonsei-Yale (Y$^{2}$) stellar evolution models
\citep{yi:2001} to determine the mass, radius, luminosity and age of
\hatcur{}. Figure~\ref{fig:iso} compares the measured values of
$T_{\rm eff}$ and $\rhostar$ to the
isochrones. Table~\ref{tab:stellar} lists the observed and derived
stellar parameters. We find that \hatcur{} has a mass of
$\hatcurISOm$\,\msun, a radius of $\hatcurISOr$\,\rsun, and is at a
reddening- and blend-corrected distance of $\hatcurXdistredblendcor{}$\,pc.

\subsection{Modeling of RVs and Light Curves}
\label{sec:globmod}

We modeled the trend-filtered HATNet light curve and the KeplerCam
light curves of \hatcur{}, together with the Keck/HIRES I$_{2}$ RVs
following the methods described by \citet{bakos:2010:hat11} and
\citet{hartman:2012:hat39hat41}. The light curves were fit using a
\citet{mandel:2002} transit model with quadratic limb darkening
coefficients fixed to the values adopted from \citet{claret:2004}. For
the KeplerCam light curves we allowed for a quadratic trend in hour
angle, and linear trends in three parameters describing the shape of
the point spread function. For the HATNet light curve we included a
dilution factor to account for distortion of the transit signal due to
the filtering procedure, and blending from neighboring stars in the
low spatial resolution HATNet images. The RVs were included in the fit
and modeled using a circular Keplerian orbit. The RV ``jitter'' term
was also varied in the fit following
\citet{hartman:2014:hat44hat46}. Note that there is no evidence that
the low cadence I$_2$ RV observations are correlated in time,
justifying the assumption of uncorrelated RV jitter. Although no
orbital variation is detected, this procedure allows us to place an
upper limit on the RV semiamplitude, and hence on the mass of
\hatcurb{}.

To account for blending in the KeplerCam light curves from \hatcur{}B
and \hatcur{}C we included contamination factors in each bandpass. We
allowed these factors to vary in the fit with Gaussian priors which
were estimated based on the measured $\Delta H$ and $\Delta
L^{\prime}$ magnitudes, together with the PARSEC isochrones, and
assuming the binary is physically associated with \hatcur{}. The
adopted priors are listed in Table~\ref{tab:mcmcparam}.

We used a Differential Evolution Markov Chain Monte Carlo (DEMCMC)
simulation \citep{terbraak:2006} to explore the likelihood function
and produce posterior distributions for all varied parameters. The parameters that we varied, together with their adopted priors, are listed in Table~\ref{tab:mcmcparam}. The
resulting Markov Chains were combined with the chains of stellar
parameters produced in Section~\ref{sec:stellar} to determine the
radius, semimajor axis, and other physical and orbital parameters for
\hatcurb{}. In particular we find that \hatcurb{} has a radius of
$R_{P} = \hatcurPPrlong{}$\,\rjup, and we place a 95\% confidence
upper limit on its mass of
$M_{P}\hatcurPPmtwosiglim$\,\mjup. Table~\ref{tab:planetparam} lists
these and other parameters for \hatcurb{}.

\subsection{Line Profile Modeling}
\label{sec:lineprof}

\begin{figure}[]
\plotone{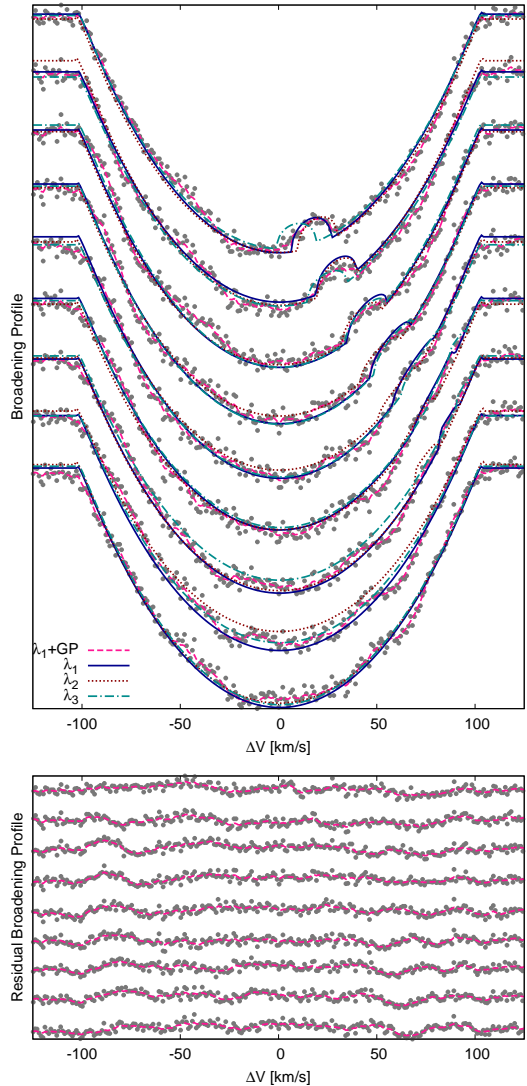}
\caption[]{
Top panel: The average rotational broadening profile of \hatcur{} for
nine consecutive Keck-I/HIRES observations proceeding chronologically
from top to bottom. The first seven profiles are from observations
obtained with \hatcurb{} in transit, while the bottom two are out of
transit. The lines labelled $\lambda_{1}$ through $\lambda_{3}$ show
models for a star undergoing solid body rotation with a quadratic limb
darkening law and a TEP with $\lambda_{1} =
\hatcurLineProflambdaone$, $\lambda_{2} = \hatcurLineProflambdatwo$
and $\lambda_{3} = \hatcurLineProflambdathree$ (in order from highest
to lowest a posteriori probability, and corresponding to the three
modes shown in Figure~\ref{fig:lambdahist}). The lines labelled
$\lambda_{1}+GP$ shows the combination of the $\lambda_{1} =
\hatcurLineProflambdaone$ physical model with a Gaussian Process
Regression used to account for additional systematic variations in the
data (see Section~\ref{sec:lineprof}). The planet creates the bump in
the profile seen at $\Delta V \sim 15$\,\kms\ in the first
observation, and progressing to higher velocities in subsequent
observations.  Bottom panel: residuals from the $\lambda_{1}$ physical
model with the Gaussian Process Regression overplotted on each. These
are displayed in the same order as in the top panel.
\label{fig:broadeningprofile}}
\end{figure}

\begin{figure}[]
\plotone{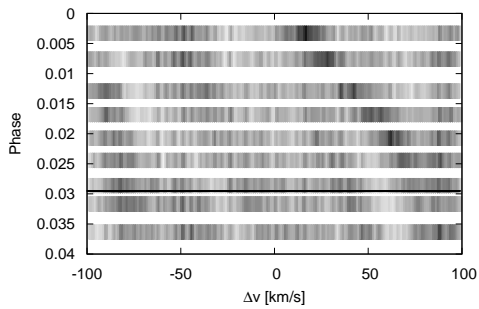}
\caption[]{
Residuals of the line profiles shown in
Figure~\ref{fig:broadeningprofile} from a simple model limb-darkened
profile, without a planet, and without including the Gaussian
Process. Zero phase corresponds to transit center, while the solid
horizontal line marks the end of transit egress. The width of each
band has been increased by a factor of 1.75 compared to the exposure
time. The greyscale has been reversed such that dark areas correspond
to positive residuals from the line profile model. The planet is seen as the
darkest shadow moving diagonally downward and to the right.
\label{fig:broadeningprofileresid}}
\end{figure}

\begin{figure}[]
\plotone{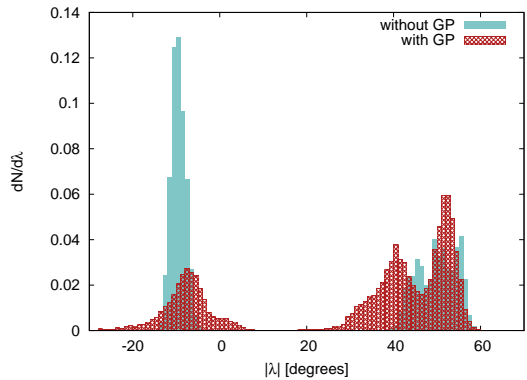}
\caption[]{
Posterior probability distribution for $\lambda$ based on modeling
the line profiles (Figure~\ref{fig:broadeningprofile}). We compare the
posterior distribution for the case when a Gaussian Process is
included to account for additional systematic variations in the line
profiles (histogram labelled ``with GP'') and for the case when a
Gaussian Process is not used (histogram labelled ``without GP''). The
posterior distribution for $\lambda$ is multimodal.
\label{fig:lambdahist}}
\end{figure}

\begin{figure*}[]
\plottwo{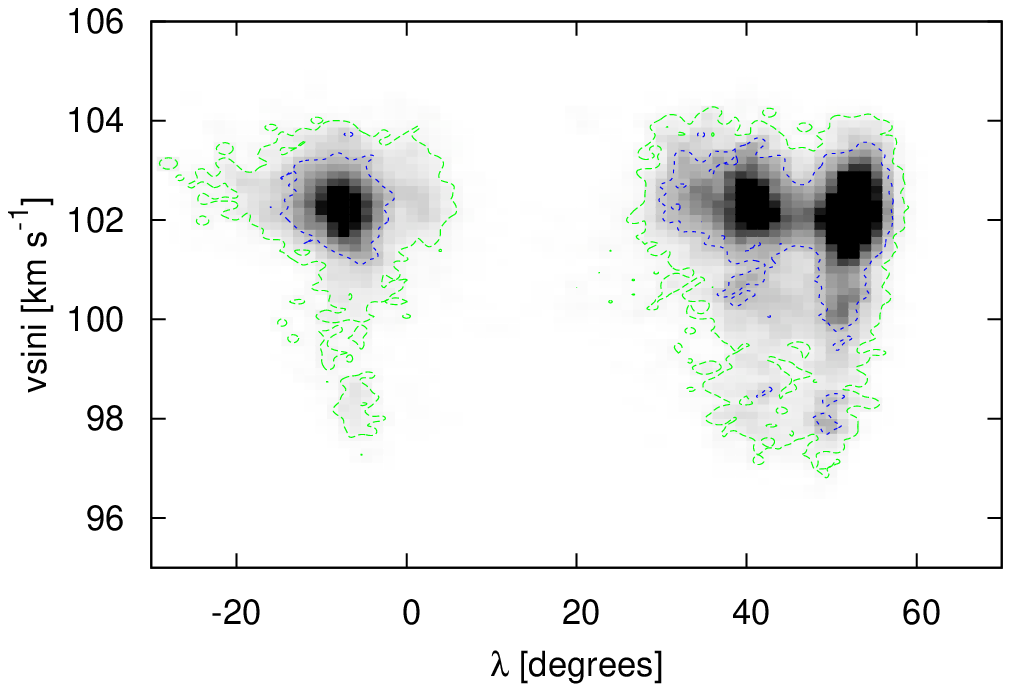}{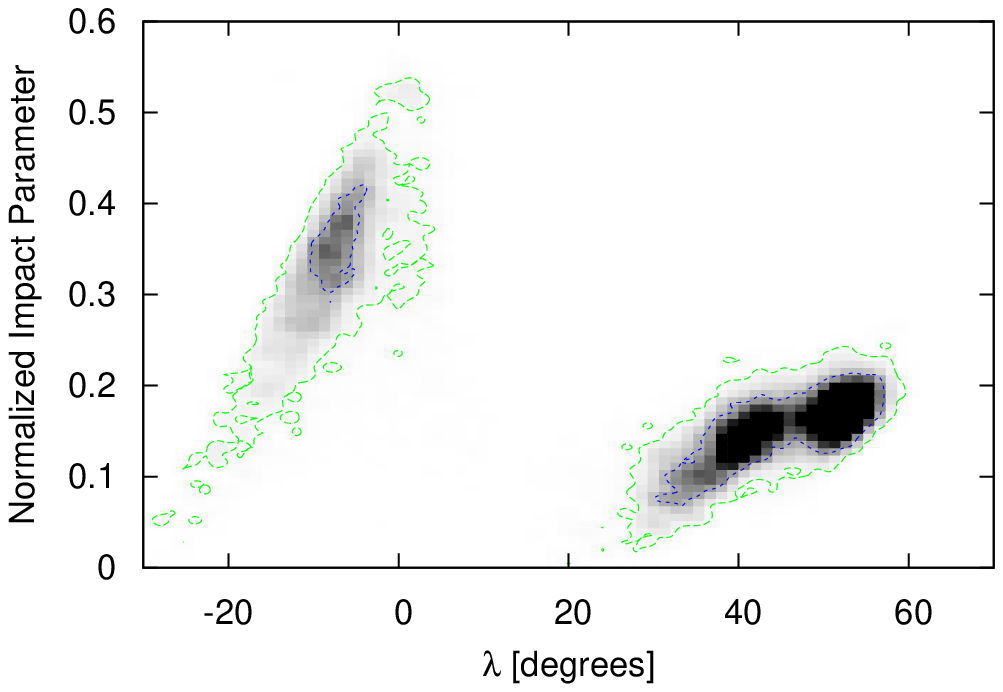}
\caption[]{
The marginalized joint posterior probability distributions for $v \sin i$ and $\lambda$ (left) and for $b$ and $\lambda$
(right) as determined from the Markov Chains produced in modeling the
line profiles (Figure~\ref{fig:broadeningprofile}). The 68.3\% and
95\% confidence contours are overplotted with blue and green lines,
respectively.
\label{fig:lambdacorr}}
\end{figure*}

\begin{figure}[]
\plotone{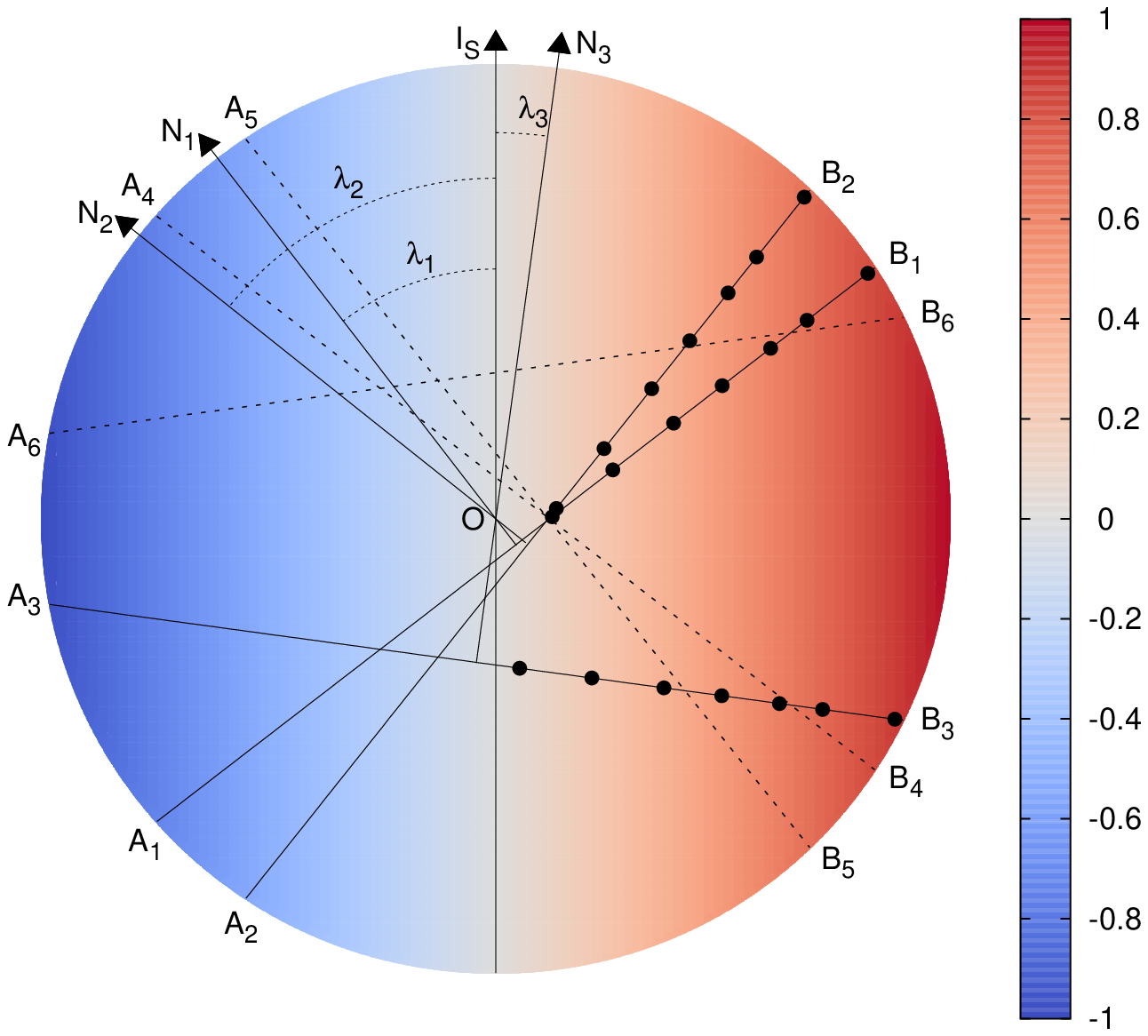}
\caption[]{
Geometry in the plane of the sky for orbital configurations of \hatcurb{} permitted by the line profile analysis. Vector $\vec{I_{S}}$ is the sky-projected spin axis of the star. The color shading indicates the rotational radial velocity at each point on the stellar surface. Transit chords $A_{1}B_{1}$, $A_{2}B_{2}$ and $A_{3}B_{3}$ are orbits with $\lambda_{1} = \hatcurLineProflambdaone$, $\lambda_{2} = \hatcurLineProflambdatwo$, and $\lambda_{3} = \hatcurLineProflambdathree$, respectively (corresponding to the three peaks in the $\lambda$ posterior distribution shown in Figure~\ref{fig:lambdahist}). These angles are measured between the spin axis of the star and the projected orbit normal vectors $\vec{N_{1}}$, $\vec{N_{2}}$, and $\vec{N_{3}}$.  Filled circles show the position of the planet at the times of the 7 Keck/HIRES observations obtained during transit. Tomography observations covering the first half of a transit would be able two distinguish between these three orbital configurations. We also show transit chords $A_{4}B_{4}$, $A_{5}B_{5}$ and $A_{6}B_{6}$ for which the tomography and light curve data are degenerate with $A_{1}B_{1}$, $A_{2}B_{2}$, and $A_{3}B_{3}$, respectively. These configurations have projected alignment angles of $180^{\circ}-\lambda$ and have the south pole of the orbit pointing toward the observer. 
\label{fig:orbitgeom}}
\end{figure}

Due to the very rapid rotation and brightness of \hatcur{}, the
in-transit Keck-I/HIRES observations are amenable to Doppler
tomography \citep[e.g.][]{colliercameron:2010b}. Because of the rapid
rotation, there are essentially no unblended lines in the spectrum of
\hatcur{}. We therefore make use of the Least Squares Deconvolution
method (\citealp{donati:1997}; see also \citealp{colliercameron:2010b}
who apply this method to observations of WASP-33) to extract the
broadening profile from the blended spectrum. Rather than using a list
of weighted delta-functions at known spectral features for the line
pattern function, as done by \citet{donati:1997}, we use the
unbroadened MARCS-atmosphere synthetic template which provided the
best match (after broadening) to the Keck-I/HIRES spectra
(Section~\ref{sec:stellar}). The deconvolution is done on an
order-by-order basis, and we then take the weighted average of 21
orders spanning 4000\AA\ to 5700\AA, excluding those containing deep
Hydrogen Balmer lines or Ca II lines.  The average broadening profiles
are shown in Figure~\ref{fig:broadeningprofile}. The residuals from a
quadratic limb-darkening rotational profile are shown in
Figure~\ref{fig:broadeningprofileresid}. The TEP is
clearly seen as the dip in Figure~\ref{fig:broadeningprofile}, or
shadow in Figure~\ref{fig:broadeningprofileresid}, moving from $\Delta
V \sim 15$\,\kms\ in the first observation near transit center to
$\Delta V \sim 77$\,\kms\ shortly before egress.

To model the broadening profile measurements we use an analytic
expression for the rotational broadening kernel of a spherical star
with quadratic limb-darkening, undergoing solid-body rotation and
transited by a spherical non-luminous planet (see
Appendix~\ref{appendix:broadeningprofile} for the derivation). This
fit was done separately from the modeling of the light curves and RVs
discussed in \refsecl{globmod}, but to ensure that the constraints on
the orbital parameters of the planet and its radius relative to the
star are incorporated into the line profile fit, we used the posterior
parameter distributions from the light curve and RV curve analysis to
determine priors on these same parameters for the line profile
fit. Our line profile model also depends on the projected angle
between the spin axis of the host and the orbital axis of the planet
($\lambda$), the maximum projected rotation velocity of the star
(\vsini), the mean velocity of the star ($\gamma$, which we take to be
a free parameter, and independent of the $\gamma$ velocity measured
with other spectrographs or reductions of the Keck-I/HIRES data), the
quadratic limb darkening coefficients ($u_{1}$ and $u_{2}$) and two
factors scaling and offsetting the model to match the
measurements. For the limb darkening coefficients we vary the
combinations $u^{\prime}_{1}$ and $u^{\prime}_{2}$ in the fit, with
$u_{1} = 0.576236u^{\prime}_{1}+0.81732928u^{\prime}_2$ and $u_{2} =
-0.81732928u^{\prime}_{1} + 0.576236u^{\prime}_{2}$, which we find to
have uncorrelated posterior probability distributions, rather than
varying $u_{1}$ and $u_{2}$ directly. The set of parameters that we
vary in this fit, together with the adopted priors, are listed in
Table~\ref{tab:mcmcparam}.

As seen in Figure~\ref{fig:broadeningprofileresid} the line profile
residuals from the physical model exhibit correlated variations that
are not associated with the planet. Based on inspecting the continuum region of the line profiles outside of the range shown in Figure~\ref{fig:broadeningprofile} we find that the systematic variations are limited to within the line profile, so we conclude that the variations are most likely due to
intrinsic stellar variability, evidence for which is also seen in the
HATNet photometry (similar variations have also been seen in the line profile of WASP-33, e.g., \citealp{colliercameron:2010b}). In order to account for these
systematic variations, which is especially important in determining
accurate uncertainties for $\lambda$ and $\vsini$, we use a non-diagonal
covariance matrix in evaluating the likelihood function. We
parameterize the covariance matrix using an exponential model, which
we found by inspection to provide a good match to the autocorrelation
of the residuals from the physical model. The covariance between points $i$ and $j$ with
velocity differences $\Delta v_{i}$ and $\Delta v_{j}$ is taken to be:
\begin{equation}
\Sigma_{ij} = \alpha\sigma_{i}^2\delta_{ij} + A\exp(-|\Delta v_{i} - \Delta v_{j}|/\rho)
\label{eqn:expcorr}
\end{equation}
where $\sigma_{i}$ is our estimated uncertainty for point $i$,
$\delta_{ij}$ is the Kronecker delta function, $\alpha$ is a parameter
used to scale the uncertainties, and $A$ and $\rho$ are parameters
describing the amplitude and length-scale for the covariance (i.e., we
are using a Gaussian-process regression, or GP, to model the systematic
variations, see \citealp{gibson:2012} for a more detailed discussion
of this technique).  The likelihood of the data in column vector ${\bf
  y}$ given the model in column vector ${\bf y_{\rm mod}}$
parameterized by $\theta_{1}$ and with covariance matrix ${\bf
  \Sigma}$ having parameters $\theta_{2}$ (i.e., $\alpha$, $A$ and
$\rho$) is then
\begin{align}
\log \mathcal{L}({\bf y}|\theta_1,\theta_2) = & -\frac{1}{2}({\bf y} - {\bf y_{\rm mod,\theta_1}})^{T}{\bf \Sigma_{\theta_2}}^{-1}({\bf y} - {\bf y_{\rm mod,\theta_1}}) \\
& - \frac{1}{2}\log |{\bf \Sigma_{\theta_2}}| + C \nonumber
\end{align}
for some constant $C$. We assume exponential priors for $\rho$ and $A$
and a Jeffreys prior for $\alpha$.

We run a DEMCMC analysis to explore this likelihood function for the line profiles, and
determine the posterior distributions of the parameters. The maximum a
posteriori model is shown in
Figure~\ref{fig:broadeningprofile}. Figure~\ref{fig:lambdahist} shows
the marginalized posterior probability distribution for $\lambda$,
while Figure~\ref{fig:lambdacorr} shows the correlations between $\lambda$ and $v \sin
i$ and between $\lambda$ and $b$. 

When the GP is used to model the systematic variations we find a
multi-modal posterior distribution for $\lambda$, with the ranges
$\hatcurLineProflambdalimonesigA^{\circ} < \lambda < \hatcurLineProflambdalimonesigB^{\circ}$,
$\hatcurLineProflambdalimonesigC^{\circ} < \lambda
<\hatcurLineProflambdalimonesigD^{\circ}$, and
$\hatcurLineProflambdalimonesigE^{\circ} < \lambda
<\hatcurLineProflambdalimonesigF^{\circ}$ having marginal posterior
probability above the 68.3\% confidence limit, and the ranges
$\hatcurLineProflambdalimtwosigA^{\circ} < \lambda < \hatcurLineProflambdalimtwosigB^{\circ}$ and
$\hatcurLineProflambdalimtwosigC^{\circ} < \lambda <
\hatcurLineProflambdalimtwosigD^{\circ}$ having marginal posterior
probability above the 95\% confidence limit. The relative
probabilities of the two modes permitted at 95\% confidence ($\hatcurLineProflambdalimtwosigA^{\circ} < \lambda <
\hatcurLineProflambdalimtwosigB^{\circ}$, and
$\hatcurLineProflambdalimtwosigC^{\circ} < \lambda <
\hatcurLineProflambdalimtwosigD^{\circ}$) are 26\% and 74\%,
respectively.

The mode peaking at $\lambda = \hatcurLineProflambdathree$ requires a
relatively high impact parameter ($b \ga 0.3$,
Figure~\ref{fig:lambdacorr}), while the higher $\lambda$ modes require
a lower impact parameter ($b \la 0.2$). This degeneracy is due in part
to the small number of line profile observations in which the planet
shadow is clearly detected, and the lack of observations prior to
transit center. Figure~\ref{fig:broadeningprofile} shows both the
overall maximum posterior probability model, which has $\lambda_{1} =
\hatcurLineProflambdaone$ and the other two local maxima ($\lambda_{2}
= \hatcurLineProflambdatwo$ and $\lambda_{3} =
\hatcurLineProflambdathree$), overplotted on the observed line
profiles, while Figure~\ref{fig:orbitgeom} shows the projected
geometry for these different orbital configurations. The models yield
similar tracks for the planet in velocity space over the time-span of
the observations. A longer time-base covering the full transit would
allow the track of the planet to be determined, and not just its
position in velocity space near transit center, helping to distinguish between
these modes. Additionally, higher precision photometric follow-up to
provide a tighter constraint on $b$ could also help break the
degeneracy. Note, however, that as seen in
Figure~\ref{fig:orbitgeom}, based solely on Doppler tomography and
transit observations, orbits with projected alignment $\lambda$ and
the north pole of the orbit pointing toward the observer are
degenerate with orbits having projected alignment $180^{\circ}-\lambda$ and the
south pole of the orbit pointing toward the observer \citep{fabrycky:2009}.

The distributions for $v \sin i$, the limb darkening coefficients, and
the correlation length scale $\rho$ are all nearly Gaussian. In
particular we find $v \sin i = \hatcurLineProfvsini$\,\kms\ and $\rho
= \hatcurLineProfrho$\,\kms. 

For the limb darkening coefficients we find $u^{\prime}_{1} =
\hatcurLineProfuA$ and $u^{\prime}_{2} = \hatcurLineProfuB$, which
correspond to $u_{1} = \hatcurLineProfuAstd$ and $u_{2} =
\hatcurLineProfuBstd$. These differ significantly from the expected
coefficients for the $g$-band from \citet{claret:2004} for a star with
the adopted atmospheric parameters of \hatcur{}. The expected values
are $u_{1,g} = \hatcurLBig$ and $u_{2,g} = \hatcurLBiig$,
respectively. Practically speaking, the \citet{claret:2004}
coefficents predict a flatter profile in the center of the line and a
steeper profile near the edge than what is observed. While errors in
the model limb darkening coefficients have been noted based on, for
example, {\em Kepler} transit observations
\citep[e.g.,][]{espinoza:2015}, the inferred $u_{1}$ coefficient based
on our observations is much larger than other works have
suggested. For example, for Kepler-13A, \citet{muller:2013} find
$u_{1} = 0.308 \pm 0.007$ and $u_{2} = 0.222 \pm 0.014$ in the {\em
  Kepler} band-pass. While models predict $u_{1}$ to be higher in the
bluer band-pass used in the Doppler tomography analysis, the expected
difference is much less than what we measure. One possible explanation
for the discrepancy is that a low order pulsation mode is distorting
the overall line profile shape, leading to incorrect limb darkening
estimates. To determine how systematic errors in the limb darkening
affect our results, we have also carried out a fit with the quadratic
limb darkening coefficients fixed to the \citet{claret:2004} $g$-band
values. We find that the posterior distribution for $\lambda$ is not
significantly affected by the treatment of limb darkening. For $v \sin
i$, on the other hand, we find that the value is more tightly
constrained when the limb darkening coefficients are fixed, but that
it is still consistent with the results when the coefficients are
allowed to vary ($v \sin i = 101.69 \pm 0.37$\,\kms\ when the
coefficients are fixed, compared to $v \sin i =
\hatcurLineProfvsini$\,\kms\ when they are allowed to vary).

For comparison, if we do not include the GP in the modeling, and
instead assume zero covariance between points in the line spread
function, then we find a bimodal, but much more tightly
constrained, distribution for $\lambda$, with the ranges $\hatcurLineProflambdanoGPlimonesigA^{\circ} < \lambda
< \hatcurLineProflambdanoGPlimonesigB^{\circ}$ and $\hatcurLineProflambdanoGPlimonesigC^{\circ} < \lambda < \hatcurLineProflambdanoGPlimonesigD^{\circ}$ having
marginal posterior probability above the 68.3\% confidence limit, and
the ranges $\hatcurLineProflambdanoGPlimtwosigA ^{\circ} < \lambda <\hatcurLineProflambdanoGPlimtwosigB ^{\circ}$ and $\hatcurLineProflambdanoGPlimtwosigC^{\circ} < \lambda <
\hatcurLineProflambdanoGPlimtwosigD^{\circ}$ having marginal posterior probability above the 95\%
confidence limit. In this case the two modes have relative
probabilities of 55\% and 45\%, respectively. The constraint on $v
\sin i$ is also tighter with $v \sin i =
\hatcurLineProfvsininoGP$\,\kms.

Other methods for modeling the systematic variations in the line
profiles were also considered. These include using a squared
exponential kernel, and using a Fourier series, but we found that the
former did not sufficiently account for long range correlations, while
the latter had the undesirable effect of suppressing both the overall
line shape and the planet signature. It is likely that a more
physically motivated model for the systematics that accounts for the
wavelike oscillations over the surface of the star, and their
propagation in time, may provide a better description of the data, and
reduce the uncertainty on $\lambda$ that results from suppressing the
planet signature through over-fitting with the GP. For example, the
two-dimensional Fourier filtering technique used by
\citet{johnson:2015} in their analysis of WASP-33 may work better at
removing the stellar oscillations. The analysis in that case was
facilitated by the retrograde motion of WASP-33, leading to a clean
separation of the planet shadow and stellar oscillations in Fourier
space. In the case of \hatcur{} the planet is prograde with $\lambda$
close to zero, and thus difficult to separate from the stellar
oscillations.

\subsection{Photometric Blend Analysis}
\label{sec:blend}

The detection of the Doppler shadow of \hatcurb{} moving across the
rotational broadening profile of \hatcur{} during a transit, and its
consistency in amplitude and width with the $\rpl/\rstar$ value
measured from the transit light curves, provides confirmation that the
transiting object is orbiting the bright rapidly rotating A star whose
light dominates the spectrum. This, coupled with the upper limit of $K
\hatcurRVKtwosiglim$\,\ms\ on the RV variation of this star, allows us
to confirm that this is a TEP system, and not a blended
stellar eclipsing binary system.

As an additional check on our conclusion that \hatcur{} is not a
blended stellar eclipsing binary system, we also carried out a blend
analysis following \citet{hartman:2012:hat39hat41}. We find that,
based on the photometry alone, all blended stellar eclipsing binary
models that we tested provide a fit to the data that has a higher
$\chi^2$ than the best-fit star+planet model. All of these blend
models can be rejected with greater than $5\sigma$ confidence in favor
of the star+planet model. Moreover, based on the MMT/Clio2 imaging,
any unaccounted-for blending companion bright enough to influence the
derived parameters of the system must be within $\sim 0\farcs25$ of
\hatcur{} (Figure~\ref{fig:contrastcurve}).

\ifthenelse{\boolean{emulateapj}}{
  \begin{deluxetable*}{lcr}
}{
  \begin{deluxetable}{lcr}
}
\tablewidth{0pc}
\tabletypesize{\scriptsize}
\tablecaption{
    Stellar Parameters for \hatcur{} 
    \label{tab:stellar}
}
\tablehead{
    \multicolumn{1}{c}{~~~~~~~~Parameter~~~~~~~~} &
    \multicolumn{1}{c}{Value}                     &
    \multicolumn{1}{c}{Source}    
}
\startdata
\noalign{\vskip -3pt}
\sidehead{Identifying Information}
~~~~R.A. (h:m:s)                      &  \hatcurCCra{} & 2MASS\\
~~~~Dec. (d:m:s)                      &  \hatcurCCdec{} & 2MASS\\
~~~~GSC ID                            &  \hatcurCCgsc{} & GSC\\
~~~~2MASS ID                          &  \hatcurCCtwomass{} & 2MASS\\
~~~~SWASP ID                          &  \hatcurCCswasp{}   & SWASP\\
\sidehead{Spectroscopic properties}
~~~~$\teffstar$ (K)\dotfill         &  \hatcurSMEteff{} & HIRES+Pollux \tablenotemark{a}\\
~~~~$\feh$\dotfill                  &  \hatcurSMEzfeh{} & HIRES+Pollux                 \\
~~~~$\vsini$ (\kms)\dotfill         &  \hatcurLineProfvsini{} & HIRES \tablenotemark{b}                 \\
~~~~$\gamma_{\rm RV}$ (\kms)\dotfill&  $-5.99 \pm 0.35$ & FIES                  \\
\sidehead{Photometric properties}
~~~~$B$ (mag)\dotfill               &  \hatcurCCtassmB{} & APASS (via URAT1)                \\
~~~~$V$ (mag)\dotfill               &  \hatcurCCtassmv{} & APASS               \\
~~~~$I$ (mag)\dotfill               &  \hatcurCCtassmI{} & TASS Mark IV   \\    
~~~~$g$ (mag)\dotfill               &  \hatcurCCtassmg{} & APASS                \\
~~~~$r$ (mag)\dotfill               &  \hatcurCCtassmr{} & APASS                \\
~~~~$i$ (mag)\dotfill               &  \hatcurCCtassmi{} & APASS                \\
~~~~$J$ (mag)\dotfill               &  \hatcurCCtwomassJmag{} & 2MASS           \\
~~~~$H$ (mag)\dotfill               &  \hatcurCCtwomassHmag{} & 2MASS           \\
~~~~$K_s$ (mag)\dotfill             &  \hatcurCCtwomassKmag{} & 2MASS           \\
\sidehead{Derived properties}
~~~~$\mstar$ ($\msun$)\dotfill      &  \hatcurISOmlong{} & Isochrones+\hatcurlumind{}+HIRES+Pollux \tablenotemark{c}\\
~~~~$\rstar$ ($\rsun$)\dotfill      &  \hatcurISOrlong{} & Isochrones+\hatcurlumind{}+HIRES+Pollux         \\
~~~~$\rhostar$ (cgs)\dotfill       &  \hatcurISOrho{} & Light Curves         \\
~~~~$\loggstar$ (cgs)\dotfill       &  \hatcurISOlogg{} & Isochrones+\hatcurlumind{}+HIRES+Pollux         \\
~~~~$\lstar$ ($\lsun$)\dotfill      &  \hatcurISOlum{} & Isochrones+\hatcurlumind{}+HIRES+Pollux         \\
~~~~$M_V$ (mag)\dotfill             &  \hatcurISOmv{} & Isochrones+\hatcurlumind{}+HIRES+Pollux         \\
~~~~$M_K$ (mag,\hatcurjhkfilset{})&  \hatcurISOMK{} & Isochrones+\hatcurlumind{}+HIRES+Pollux         \\
~~~~Age (Gyr)\dotfill               &  \hatcurISOage{} & Isochrones+\hatcurlumind{}+HIRES+Pollux         \\
~~~~$A_{V}$ (mag) \tablenotemark{d}\dotfill           &  \hatcurXAvblendcor{} & Isochrones+\hatcurlumind{}+HIRES+Pollux\\
~~~~Distance (pc) \tablenotemark{e}\dotfill           &  \hatcurXdistredblendcor{} & Isochrones+\hatcurlumind{}+HIRES+Pollux\\
\enddata
\tablenotetext{a}{
    HIRES+Pollux = Based on a $\chi^2$ comparison between the
    extracted HIRES spectra and synthetics MARCS model atmosphere
    spectra \citep{gustafsson:2008} from the Pollux database \citep{palacios:2010} as
    discussed in Section~\ref{sec:stellar}.
}  
\tablenotetext{b}{
    Based on modeling the spectral line profiles as discussed in Section~\ref{sec:lineprof}.
}
\tablenotetext{c}{
    Isochrones+\hatcurlumind{}+HIRES+Pollux = Based on the Y$^{2}$ isochrones
    \citep{yi:2001},
    the stellar density used as a luminosity indicator, and the atmospheric parameter results.
} 
\tablenotetext{d}{ Total \band{V} extinction to the star determined
  by comparing the catalog broad-band photometry listed in the table
  to the expected magnitudes from the
  Isochrones+\hatcurlumind{}+HIRES+Pollux model for the star, and accounting for blending from the known binary located $2\farcs7$ away from \hatcur{}. We use the
  \citet{cardelli:1989} extinction law.  }
\tablenotetext{e}{ Distance based on a comparison of the measured photometric magnitudes for \hatcur{}, corrected for blending from \hatcur{}A and \hatcur{}B and for reddening, to the predicted magnitudes from the stellar evolution models.}
\ifthenelse{\boolean{emulateapj}}{
  \end{deluxetable*}
}{
  \end{deluxetable}
}

\ifthenelse{\boolean{emulateapj}}{
  \begin{deluxetable}{ll}
}{
  \begin{deluxetable}{ll}
}
\tabletypesize{\scriptsize}
\tablecaption{MCMC State Variables and Priors.\label{tab:mcmcparam}}
\tablehead{
    \multicolumn{1}{c}{~~~~~~~~Parameter~~~~~~~~} &
    \multicolumn{1}{c}{Prior}
}
\startdata
\noalign{\vskip -3pt}
\sidehead{\Lc{}+RV curve analysis}
~~~$T_{c,0}$ (days) \tablenotemark{a}            \dotfill    & uniform    \\
~~~$T_{c,888}$ (days) \tablenotemark{a}            \dotfill    & uniform    \\
~~~$\zrstar$              \dotfill    & uniform \\
~~~$\rpl/\rstar$          \dotfill    & uniform       \\
~~~$b^2$                  \dotfill    & uniform  with $0 \leq b^{2} \leq 1$          \\
~~~$K$ (\kms) \dotfill & uniform with $K \geq 0$ \\
~~~$\gamma_{\rm rel}$ HIRES \dotfill & uniform \\
~~~RV jitter HIRES & uniform with $\sigma_{\rm jitter} \geq 0$ \\
~~~$f_{\rm blend, HN, r}$ \tablenotemark{b} \dotfill & uniform with $0 \leq f_{\rm blend, HN} \leq 1$ \\
~~~$f_{\rm blend, g}$ \tablenotemark{b} \dotfill & N(0.9980,0.0026) \tablenotemark{c} with $0 \leq f_{\rm blend, g} \leq 1$ \\
~~~$f_{\rm blend, i}$ \tablenotemark{b} \dotfill & N(0.9836,0.0088) \tablenotemark{c} with $0 \leq f_{\rm blend, i} \leq 1$ \\
~~~$f_{\rm blend, z}$ \tablenotemark{b} \dotfill & N(0.973,0.011) \tablenotemark{c} with $0 \leq f_{\rm blend, z} \leq 1$ \\
~~~$m_{0}$ \tablenotemark{d} \dotfill & uniform, linearly optimized \\
~~~$c_{EPD}$ \tablenotemark{e} \dotfill & uniform, linearly optimized \\
\noalign{\vskip -3pt}
\sidehead{Line profile analysis}
~~~$T_{c}$ (days)             \dotfill    & N(2455113.48127,0.00048) \tablenotemark{c} \\
~~~$P$ (days)             \dotfill    & N(2.4652950,0.0000032) \tablenotemark{c} \\
~~~$\arstar$              \dotfill    & N(5.825,0.090) \tablenotemark{c} \\
~~~$\rpl/\rstar$          \dotfill    & N(0.0968,0.0015) \tablenotemark{c} \\
~~~$b^2$                  \dotfill    & N(0.051,0.023) \tablenotemark{c} with $0 \leq b^2 \leq 1$ \\
~~~$\lambda$  (days)      \dotfill    & uniform $-180 \leq \lambda < 180$ \\
~~~$v \sin i$ (\kms)      \dotfill    & uniform \\
~~~$\Delta v_{0}$ (\kms) \tablenotemark{f}  \dotfill    & uniform \\
~~~$u_{1}^{\prime}$         \dotfill    & uniform subject to $0 \leq u_{1} \leq 1$ \\
~~~$u_{2}^{\prime}$         \dotfill    & uniform subject to $0 \leq u_{2} \leq 1$\\
~~~$\rho$ \tablenotemark{g} \dotfill & $\propto e^{-\rho/200}$ with $0 \leq \rho \leq 200$ \\
~~~$A$ \tablenotemark{g} \dotfill & $\propto e^{-A/0.0000002}$ \\
~~~$\alpha$ \tablenotemark{g} \dotfill & $\propto 1/\alpha$ with $\alpha > 0$ \\
~~~$a_{\rm LP}$ \tablenotemark{h} \dotfill & uniform \\
~~~$c_{\rm LP}$ \tablenotemark{i} \dotfill & uniform \\
\enddata
\tablenotetext{a}{
    The times of transit center for event number $0$ (the first transit covered by our light curves) and event number $888$ (the last transit covered by our light curves).
}
\tablenotetext{b}{
    Scaling factors for each filter applied to the fractional stellar
    flux blocked by the planet to account for dilution from \hatcur{}B
    and \hatcur{}C in all of the light curves, and to account for
    over-filtering in the HATNet data.
}
\tablenotetext{c}{
    Here $N(\mu,\sigma)$ corresponds to a normal distribution with
    mean $\mu$ and standard deviation $\sigma$. For the blend factors
    these are determined based on the measured magnitudes of
    \hatcur{}B and \hatcur{}C together with the PARSEC isochrones. For the
    line profile parameters these are determined from the posterior
    distributions for each parameter from the light curve and RV curve
    analysis.
}
\tablenotetext{d}{
    Out-of-transit magnitude. One such parameter is used for each
    light curve in the analysis. For computational efficiency these
    parameters are optimized via linear least squares at each step in
    the MCMC.
}
\tablenotetext{e}{
    EPD coefficients used to remove quadratic variations in time, or
    variations that are correlated with changes in the shape of the
    PSF.  Five such parameters are used for each of the KeplerCam
    light curves. For computational efficiency these parameters are
    optimized via linear least squares at each step in the MCMC.
}
\tablenotetext{f}{
    Center velocity of the line profile. One such parameter is used
    for each profile analyzed.
}
\tablenotetext{g}{
    Noise model parameters discussed in Section~\ref{sec:lineprof}.
}
\tablenotetext{h}{
    Parameter scaling the depth of the line profile. One such
    parameter is used for each profile analyzed.
}
\tablenotetext{i}{
    The continuum level of the line profile. One such parameter is
    used for each profile analyzed.
}
\ifthenelse{\boolean{emulateapj}}{
  \end{deluxetable}
}{
  \end{deluxetable}
}
%

\ifthenelse{\boolean{emulateapj}}{
  \begin{deluxetable*}{lclc}
}{
  \begin{deluxetable}{lclc}
}
\tabletypesize{\scriptsize}
\tablecaption{Parameters for the transiting planet \hatcur{}${\mathrm{b}}$.\label{tab:planetparam}}
\tablehead{
    \multicolumn{1}{c}{~~~~~~~~Parameter~~~~~~~~} &
    \multicolumn{1}{c}{Value \tablenotemark{a}} &
    \multicolumn{1}{c}{~~~~~~~~Parameter~~~~~~~~} &
    \multicolumn{1}{c}{Value \tablenotemark{a}}
}
\startdata
\noalign{\vskip -3pt}
\sidehead{\Lc{} parameters}
~~~$P$ (days)             \dotfill    & $\hatcurLCP{}$    &
~~~$T_c$ (${\rm BJD}$)    
      \tablenotemark{b}   \dotfill    & $\hatcurLCT{}$              \\
~~~$T_{14}$ (days)
      \tablenotemark{b}   \dotfill    & $\hatcurLCdur{}$   &
~~~$T_{12} = T_{34}$ (days)
      \tablenotemark{b}   \dotfill    & $\hatcurLCingdur{}$         \\
~~~$\arstar$              \dotfill    & $\hatcurPPar{}$           &
~~~$\zrstar$ \tablenotemark{c}              \dotfill    & $\hatcurLCzeta{}$\phn       \\
~~~$\rpl/\rstar$          \dotfill    & $\hatcurLCrprstar{}$       &
~~~$b^2$                  \dotfill    & $\hatcurLCbsq{}$            \\
~~~$b \equiv a \cos i/\rstar$
                          \dotfill    & $\hatcurLCimp{}$          &
~~~$i$ (deg)              \dotfill    & $\hatcurPPi{}$\phn         \\

\sidehead{Line Profile parameters}
~~~$\lambda$ (deg) \tablenotemark{d}     \dotfill    & $\hatcurLineProflambdalimtwosigA < \lambda < \hatcurLineProflambdalimtwosigB$ or $\hatcurLineProflambdalimtwosigC < \lambda < \hatcurLineProflambdalimtwosigD$ &
~~~$\vsini$ (\kms)        \dotfill    & $\hatcurLineProfvsini$           \\
~~~$\rho$ (\kms) \tablenotemark{e}        \dotfill    & $\hatcurLineProfrho$           &
~~~$u^{\prime}_{1}$ \tablenotemark{f}       \dotfill    & $\hatcurLineProfuA$           \\
~~~$u^{\prime}_{2}$ \tablenotemark{f}       \dotfill    & $\hatcurLineProfuB$           & 
~~~$u_{1}$ \dotfill & $\hatcurLineProfuAstd$ 
     \\
~~~$u_{2}$ \dotfill & $\hatcurLineProfuBstd$
      &  & \\

\sidehead{Limb-darkening coefficients \tablenotemark{g}}
~~~$c_1,g$ (linear term)  \dotfill    & $\hatcurLBig{}$            &
~~~$c_2,g$ (quadratic term) \dotfill  & $\hatcurLBiig{}$           \\
~~~$c_1,i$               \dotfill    & $\hatcurLBii{}$            &
~~~$c_2,i$               \dotfill  & $\hatcurLBiii{}$           \\
~~~$c_1,r$               \dotfill    & $\hatcurLBir{}$             &
~~~$c_2,r$               \dotfill    & $\hatcurLBiir{}$            \\
~~~$c_1,z$               \dotfill    & $\hatcurLBiz{}$             &
~~~$c_2,z$               \dotfill    & $\hatcurLBiiz{}$            \\

\sidehead{RV parameters}
~~~$K$ (\ms)  \tablenotemark{h}            \dotfill    & $\hatcurRVKtwosiglim{}$\phn\phn      &
~~~$e$ \tablenotemark{i}  \dotfill    & $0$ \\
~~~RV jitter (\ms) \tablenotemark{j}        \dotfill    & \hatcurRVjitter{}           & & \\

\sidehead{Planetary parameters}
~~~$\mpl$ ($\mjup$) \tablenotemark{h}      \dotfill    & $\hatcurPPmtwosiglim{}$          &
~~~$\rpl$ ($\rjup$)       \dotfill    & $\hatcurPPrlong{}$          \\
~~~$a$ (AU)               \dotfill    & $\hatcurPParel{}$          &
~~~$T_{\rm eq}$ (K) \tablenotemark{k}        \dotfill   & $\hatcurPPteff{}$           \\
~~~$\langle F \rangle$ ($10^{\hatcurPPfluxavgdim{}}$\ergscmsq) \tablenotemark{l}
                          \dotfill    & $\hatcurPPfluxavg{}$     & &  \\ [-1.5ex]
\enddata
\tablenotetext{a}{
    The adopted parameters assume a circular orbit. Based on the
    Bayesian evidence ratio we find that this model is strongly
    preferred over a model in which the eccentricity is allowed to
    vary in the fit. For each parameter we give the median value and
    68.3\% (1$\sigma$) confidence intervals from the posterior
    distribution.
}
\tablenotetext{b}{
    Reported times are in Barycentric Julian Date calculated directly
    from UTC, {\em without} correction for leap seconds.
    \ensuremath{T_c}: Reference epoch of mid transit that
    minimizes the correlation with the orbital period.
    \ensuremath{T_{14}}: total transit duration, time
    between first to last contact;
    \ensuremath{T_{12}=T_{34}}: ingress/egress time, time between first
    and second, or third and fourth contact.
}
\tablenotetext{c}{
    Reciprocal of the half duration of the transit used as a jump
    parameter in our MCMC analysis in place of $\arstar$. It is
    related to $\arstar$ by the expression $\zrstar = \arstar
    (2\pi(1+e\sin \omega))/(P \sqrt{1 - b^{2}}\sqrt{1-e^{2}})$
    \citep{bakos:2010:hat11}.
}
\tablenotetext{d}{
    The marginalized posterior probability distribution for $\lambda$ is multimodal. We list the ranges of $\lambda$ above the 95\% confidence level. The two ranges have relative probabilities of 26\% and 74\%, respectively.
}
\tablenotetext{e}{
    The exponential correlation length scale for describing systematic
    variations in the line profile (Eq.~\ref{eqn:expcorr}).
}
\tablenotetext{f}{
    Uncorrelated quadratic limb darkening coefficients from modeling
    the spectral line profiles. These are related to the usual
    quadratic limb coefficients $u_{1}$ and $u_{2}$ via $u_{1} =
    0.576236u^{\prime}_{1}+0.81732928u^{\prime}_2$ and $u_{2} =
    -0.81732928u^{\prime}_{1} + 0.576236u^{\prime}_{2}$.
}
\tablenotetext{g}{
    Values for a quadratic law, adopted from the tabulations by
    \cite{claret:2004} according to the spectroscopic (SPC) parameters
    listed in \reftabl{stellar}.
}
\tablenotetext{h}{
    95\% confidence upper limit.
}
\tablenotetext{i}{
    We assume a circular orbit for the analysis.
}
\tablenotetext{j}{
    Error term, either astrophysical or instrumental in origin, added
    in quadrature to the formal RV errors. This term is varied in the
    fit assuming a prior inversely proportional to the jitter.
}
\tablenotetext{k}{
    Planet equilibrium temperature averaged over the orbit, calculated
    assuming a Bond albedo of zero, and that flux is reradiated from
    the full planet surface.
}
\tablenotetext{l}{
    Incoming flux per unit surface area, averaged over the orbit.
}
\ifthenelse{\boolean{emulateapj}}{
  \end{deluxetable*}
}{
  \end{deluxetable}
}
%



\section{Discussion}
\label{sec:discussion}

In this paper we have presented the discovery of \hatcurb{}, a
short-period ($P = \hatcurLCPshort$\,days) giant planet transiting a
rapidly rotating A8V star. Periodic photometric transits in the light
curve of this source have been independently detected by three
separate transit surveys (HAT, WASP and KELT; see
Section~\ref{sec:detection}). Here we combine the HATNet photometry,
with follow-up photometry from FLWO~1.2\,m/KeplerCam and spectroscopy
from Keck-I/HIRES, NOT/FIES and FLWO~1.5\,m/TRES to confirm that this
is a transiting planet system and to determine its properties. The
confirmation follows from three pieces of observational evidence: (1)
Keck-I/HIRES spectroscopy obtained during a transit reveals the
Doppler shadow of the planet \hatcurb{} moving across the average
spectral absorption line profile of the star \hatcur{}. The
consistency of the shape of the shadow with the transiting planet
parameters measured from the light curve proves that the transiting
object is not orbiting a fainter object blended by the bright A star
whose light dominates the spectrum. (2) Keck-I/HIRES RVs obtained out
of transit allow us to place an upper limit on the semiamplitude of
the RV orbital variation of the A star of $K\hatcurRVKtwosiglim$\,\ms,
and a corresponding upper limit on the mass of the transiting object
of $M\hatcurPPmtwosiglim$\,\mjup. (3) A blend analysis of the
available photometric data rules out blended eclipsing binary
scenarios in favor of a single star with a transiting planet with
greater than 5$\sigma$ confidence.

Based on our analysis of the photometric and spectroscopic data,
together with the Y$^{2}$ stellar evolution models, we conclude that the
star \hatcur{} has a mass of \hatcurISOm\,\msun, a radius of
\hatcurISOr\,\rsun, and is located at a distance of
\hatcurXdistredblendcor\,pc from the Solar System. The planet
\hatcurb{} has a radius of \hatcurPPr\,\rjup, a semimajor axis of
\hatcurPParel\,AU, and an estimated equilibrium temperature (assuming
zero albedo and complete redistribution of heat) of \hatcurPPteff\,K.

Adaptive optics imaging in $H$ and $L^{\prime}$ bands performed with
MMT/Clio2 reveals a pair of stars separated
$\hatcurXCompanionOuterSepshort$ from \hatcur{} and
$\hatcurXCompanionInnerSepshort$ from each other. The stars have $H$
and $L^{\prime}$ magnitudes consistent with being stars of mass
\hatcurXCompanionStarMassPARSECA\,\msun\ and
\hatcurXCompanionStarMassPARSECB\,\msun\ located at the same distance
from the Solar System as \hatcur{}. If they are physically associated
with \hatcur{}, then this is a hierarchical triple star system with
\hatcur{}B and \hatcur{}C having a projected physical separation of
$\hatcurXCompanionInnerPhysSeplong$\,AU, and approximate orbital
period of $\hatcurXCompanionInnerPeriodApprox$\,yr (assuming the
projected separation corresponds to the physical semimajor axis of the
orbit), while the \hatcur{}B+\hatcur{}C binary has a projected
physical separation of $\hatcurXCompanionOuterPhysSeplong$\,AU from
\hatcur{} and an approximate orbital period of
$\hatcurXCompanionOuterPeriodApprox$\,yr. 

There are two factors which distinguish \hatcurb{} from the more than
$1200$ other confirmed or validated transiting planet systems. With a
projected equatorial rotation velocity of $\vsini =
\hatcurLineProfvsini$\,\kms, \hatcur{} has the highest rotation
velocity of any star known to host a transiting planet. The next most
rapidly rotating stars with transiting planets are KOI-89 ($v \sin i \approx 90$\,\kms, \citealp{ahlers:2015}), WASP-33 ($v \sin i
= 86.48 \pm 0.06$\,\kms, \citealp{colliercameron:2010b}), Kepler-13A
($v \sin i = 76.6 \pm 0.2$\,\kms, \citealp{santerne:2012}), and KELT-7
($v \sin i = 65 \pm 6$\,\kms, \citealp{bieryla:2015}). \hatcur{},
together with WASP-33, Kepler-13A, and KOI-89, are also the only four stars
of spectral type earlier than F0 known to host transiting
planets. This makes \hatcurb{} a valuable system for studying the
properties of close-in planets around rapidly rotating, relatively
high mass stars.

While the rapid rotation of \hatcur{} prevents us from measuring the
mass of the planet through the RV orbital wobble of the host star, it
also creates an opportunity to characterize the orbital geometry of
the planetary system with unusually high accuracy. We have already
taken initial steps in this direction through our modeling of the
spectral line profiles during a partial transit event observed with
Keck-I/HIRES. We constrain the projected angle between the spin axis
of the host star and the orbital axis of the planet to lie within the
range $\hatcurLineProflambdalimtwosigA^{\circ} < \lambda < \hatcurLineProflambdalimtwosigB^{\circ}$ or
$\hatcurLineProflambdalimtwosigC^{\circ} < \lambda <
\hatcurLineProflambdalimtwosigD^{\circ}$ with 95\% confidence. These
two distinct modes have relative probabilities of 26\% and 74\%,
respectively. While we do not find a unique solution, we are able to
rule out very high obliquities, and conclude that \hatcurb{} is either
moderately misaligned in projection (most likely) or it is close to being aligned in projection (less likely, but still possible). Additional
Doppler tomography observations, especially observations
covering a full transit, will be able to distinguish between these two
scenarios and pin down the angle $\lambda$.

If \hatcurb{} is not on a well-aligned orbit, then the significant
expected oblateness for \hatcur{}, resulting from its rapid rotation,
should cause the orbit of \hatcurb{} to precess at a relatively rapid
rate. This would be observable by measuring changes in $\lambda$
and/or $b$ over time, as has been done for WASP-33b
\citep{johnson:2015} and Kepler-13Ab \citep{szabo:2012,
  masuda:2015}. Measuring the precession rate would provide an
observational constraint on the $J_{2}$ gravitational quadrupole
moment of the star, which in turn may lead to a better age
determination for the system, and/or can be used to test the
theoretical stellar evolution models themselves. 

The rapid rotation of \hatcur{} also makes possible a measurement of
the true (not projected) spin--orbit alignment angle by detecting an
asymmetry in the transit shape resulting from gravity darkening in the
oblate star \citep{masuda:2015} (the degeneracy between orbits having
projected alignment angle $180^{\circ}-\lambda$ seen in
Figure~\ref{fig:orbitgeom} may also be lifted by this technique). Following
\citet{vanbelle:2004}, we estimate that \hatcur{} has an oblateness of
$R_{b}/R_{a} \approx 1.03$, where $R_{b}$ and $R_{a}$ are the
equatorial and polar radii of the star, respectively. This is quite
similar to the estimate for WASP-33. Following \citet{zhou:2013}, we
estimate that the maximum difference between transit models including
and excluding the gravity darkening effect is $\approx 500$\,ppm,
assuming the rotation axis of the star lies in the plane of the
sky. Detecting such a signal may be possible, though challenging, from
the ground. It should also be detectable by TESS (based on its
position on the sky, we expect TESS to monitor \hatcur{} for $\sim
27$\,days). This will depend, however, on whether \hatcur{} exhibits
high frequency (i.e., $\delta$-Scuti or roAp-type variations)
photometric oscillations with amplitudes greater than the gravity
darkening effect. While the HATNet light curve rules out high
frequency oscillations with an amplitude above 1.2\,mmag in $r$-band
(the observed low frequency oscillations are on a long enough time
scale that it should be possible to filter these from transit light
curves), the variations seen in the Keck-I/HIRES line profiles
indicate that lower amplitude high frequency oscillations may be
present.


\acknowledgements 

\paragraph{Acknowledgements}
We are grateful to the anonymous referee for their careful review of
this paper, including several important comments. HATNet operations
have been funded by NASA grants NNG04GN74G and NNX13AJ15G. Follow-up
of HATNet targets has been partially supported through NSF grant
AST-1108686. G.\'A.B., Z.C. and K.P. acknowledge partial support from
NASA grant NNX09AB29G.  J.H. acknowledges support from NASA grant
NNX14AE87G. K.P. acknowledges support from NASA grant NNX13AQ62G.  We
acknowledge partial support also from the Kepler Mission under NASA
Cooperative Agreement NCC2-1390 (D.W.L., PI). We would also like to
thank J.~Pepper for informing us about the KELT observations of this
system. Data presented in this paper are based on observations
obtained at the HAT station at the Submillimeter Array of SAO, and the
HAT station at the Fred Lawrence Whipple Observatory of SAO. The
authors wish to recognize and acknowledge the very significant
cultural role and reverence that the summit of Mauna Kea has always
had within the indigenous Hawaiian community. We are most fortunate to
have the opportunity to conduct observations from this mountain. This
paper presents observations made with the Nordic Optical Telescope,
operated on the island of La Palma jointly by Denmark, Finland,
Iceland, Norway, and Sweden, in the Spanish Observatorio del Roque de
los Muchachos of the Instituto de Astrof\'isica de Canarias.

\clearpage

\appendix

\section{Analytic Rotational Broadening Kernel For a Quadratically Limb Darkened Star With a Transiting Planet}
\label{appendix:broadeningprofile}

The Doppler tomography method has been used to determine the
spin-orbit alignments of the HD~189733, HAT-P-2, WASP-32, WASP-33,
WASP-38, Kepler-13A, Kepler-25, and KOI-12 transiting planet systems
\citep{colliercameron:2010a,colliercameron:2010b,brown:2012,albrecht:2013,johnson:2014,johnson:2015,bourrier:2015}. These
previous applications have modeled the rotational broadening function
by using an analytic model for the broadening profile of a star with
linear limb darkening and a transiting planet having $R_{P}/R_{\star}
\ll 1$ \citep[i.e., the planet shadow is treated as a
  Gaussian,][]{colliercameron:2010a,colliercameron:2010b,brown:2012,bourrier:2015},
or by carrying out a numerical integration of a gridded stellar
surface brightness profile
\citep{albrecht:2013,johnson:2014,johnson:2015}.  The rotational
broadening function for a star with linear limb darkening, undergoing
solid body rotation, and being eclipsed by another object has a simple
analytic form which was worked out by \citet{kopal:1959}. Here we
provide the analogous formula for a quadratic limb darkening law. We
caution that this relation is only applicable for large rotation rates
where macro-turbulence may be neglected.

The spectrum of a rotating star can be calculated by convolving the
non-rotating spectrum with a broadening kernel
\citep[e.g.,][]{gray:2005}, i.e.,
\begin{equation}
S_{R}(\lambda) = \int_{-\infty}^{\infty}S(\lambda(1 - \tilde{v}v_{L}/c))G(\tilde{v})d\tilde{v}
\end{equation}
where $S(\lambda)$ is the non-rotating spectrum at wavelength
$\lambda$, $v_{L}$ is the projected rotation velocity of the star, $c$
is the speed of light, and $G(\tilde{v})$ is the broadening kernel
evaluated at relative velocity shift $\tilde{v}$.  Following \citet{gray:2005}, the rotational broadening kernel is given by:
\begin{equation}
G(\tilde{v}) = \frac{G'(\tilde{v})}{\int_{-\infty}^{\infty}G'(v')dv'}
\end{equation}
with
\begin{equation}
\label{eqn:gprimedef}
G'(\tilde{v}) = \int_{y_{\rm min}(\tilde{v})}^{y_{\rm max}(\tilde{v})}I(\tilde{v},y)dy
\end{equation}
being the integral of the stellar surface brightness $I(x,y)$ along a
line of constant $x = \tilde{v}$. Here we are using a coordinate
system centered on the stellar disk with the $y$ axis parallel to
the projected rotation axis of the star, and with $x$ and $y$ measured
in units of the stellar radius (note that for solid body rotation the
projected rotation velocity is constant along a line of constant
$x$). We also have $y_{\rm min}(x) = -\sqrt{1-x^2}$ and $y_{\rm max}(x)
= \sqrt{1-x^2}$.

For a quadratic limb darkening law of the form
\begin{equation}
I(\theta) = I_{c}(1 - u_{1}(1 - \cos(\theta)) - u_{2}(1 - \cos(\theta))^2)
\end{equation}
with $\theta$ being the angle between the line normal to the stellar surface and the line of sight from the center of the star to the observer,
equation~\ref{eqn:gprimedef} works out to
\begin{equation}
\label{eqn:gprimequad}
G'(\tilde{v}) = \begin{cases}
    2(1-u_{1}-u_2)\sqrt{1 - \tilde{v}^{2}} + \frac{\pi}{2}(u_{1} + 2u_{2})(1 - \tilde{v}^2) - \frac{4}{3} u_{2}(1-\tilde{v}^2)^{3/2} & |\tilde{v}| < 1 \\
    0 & |\tilde{v}| \geq 1
\end{cases}
\end{equation}
and
\begin{equation}
\int_{-\infty}^{\infty} G'(v')dv' = \pi(1 - u_{1}/3 - u_{2}/6)
\end{equation}
When a transiting planet with radius $R$ (in units of the stellar
radius), is in front of the star at projected position $(x_{P},y_{P})$
relative to the center of the star (these are determined from the
orbital parameters and the projected spin--orbit angle $\lambda$ following,
e.g., \citealp{boue:2013}), we must subtract
\begin{equation}
K(\tilde{v}) = \int_{y_{1}(\tilde{v},x_{P},y_{P},R)}^{y_{2}(\tilde{v},x_{P},y_{P},R)}I(\tilde{v},y)dy,
\end{equation}
the integral of the stellar surface brightness blocked by the planet,
from equation~\ref{eqn:gprimequad}. For a quadratic limb darkening law we have
\begin{equation}
\begin{split}
K(\tilde{v})&= (y_{2} - y_{1})(1-u_{1}-u_{2}(2 - \tilde{v}^2))+(\frac{1}{2}u_{1} + \frac{2}{3}u_{2})(y_{1}^3 - y_{2}^3)\\
            &+\frac{1}{2}(u_{1}+2u_{2})\left[(y_{2}-y_{1})(1-\tilde{v}^{2})+(1-\tilde{v}^2)\left(\sin^{-1}\left(\frac{y_{2}}{\sqrt{1-\tilde{v}^2}}\right)-\sin^{-1}\left(\frac{y_{1}}{\sqrt{1-\tilde{v}^2}}\right)\right)\right]
\end{split}
\end{equation}
with
\begin{equation}
y_{1}(\tilde{v},x_{P},y_{P},R) = \begin{cases}
0 & |\tilde{v}| \geq 1 {\,\rm or\,} |\tilde{v}-x_{P}| \geq R \\
\sqrt{1 - \tilde{v}^2} & y_{P} - \sqrt{R^{2} - (\tilde{v}-x_{P})^2} \geq \sqrt{1-\tilde{v}^2} \\
-\sqrt{1 - \tilde{v}^2} & y_{P} - \sqrt{R^{2} - (\tilde{v}-x_{P})^2} \leq \sqrt{1-\tilde{v}^2} \\
y_{P} - \sqrt{R^{2} - (\tilde{v}-x_{P})^2} & {\rm otherwise}
\end{cases}
\end{equation}
and
\begin{equation}
y_{2}(\tilde{v},x_{P},y_{P},R) = \begin{cases}
0 & |\tilde{v}| \geq 1 {\,\rm or\,} |\tilde{v}-x_{P}| \geq R \\
\sqrt{1 - \tilde{v}^2} & y_{P} + \sqrt{R^{2} - (\tilde{v}-x_{P})^2} \geq \sqrt{1-\tilde{v}^2} \\
-\sqrt{1 - \tilde{v}^2} & y_{P} + \sqrt{R^{2} - (\tilde{v}-x_{P})^2} \leq \sqrt{1-\tilde{v}^2} \\
y_{P} + \sqrt{R^{2} - (\tilde{v}-x_{P})^2} & {\rm otherwise}
\end{cases}
\end{equation}
Examples of this model fit to the measured broadening profiles of
\hatcur{} are shown in Fig~\ref{fig:broadeningprofile}, where we have
inverted the model profile to look like absorption lines for display
purposes.

\end{document}